\documentclass[%
aip,
amsmath,amssymb,aps,
reprint,%
]{revtex4-1}

\usepackage{graphicx}
\usepackage{hyperref}
\usepackage{dcolumn}
\usepackage{color}
\usepackage{bm}

\usepackage[utf8]{inputenc}
\usepackage[T1]{fontenc}
\usepackage{mathptmx}

\newcommand{\comment}[1]{}

\begin{document}
	
	\preprint{AIP/123-QED}
	
	\title[]{Quantum Anomalous Hall Effect in $d$-Electron Kagome Systems: Chern Insulating States from Transverse Spin-Orbit Coupling}
	
	\author{Imam Makhfudz}
	
	\affiliation{IM2NP, UMR CNRS 7334, Aix-Marseille Universit\'{e}, 13013 Marseille, France}
	\author{Mikhail  Cherkasskii}
	\affiliation{Institut für Theoretische Festkörperphysik, RWTH Aachen University, 52056 Aachen, Germany}
	\author{Mohammad Alipourzadeh}
	\affiliation{Department of Physics, Faculty of Science, Shahid Chamran University of Ahvaz, 6135743135 Ahvaz, Iran}
	\author{Yaser Hajati}
	\affiliation{Department of Physics, Faculty of Science, Shahid Chamran University of Ahvaz, 6135743135 Ahvaz, Iran}
	\author{Pierre Lombardo}
	\affiliation{IM2NP, UMR CNRS 7334, Aix-Marseille Universit\'{e}, 13013 Marseille, France}
	\author{Steffen Schäfer}
	\affiliation{IM2NP, UMR CNRS 7334, Aix-Marseille Universit\'{e}, 13013 Marseille, France}
	\author{Silvia Viola Kusminskiy}
	\affiliation{Institut für Theoretische Festkörperphysik, RWTH Aachen University, 52056 Aachen, Germany}
	\affiliation{Max Planck Institute for the Science of Light, Staudtstraße 2, PLZ 91058 Erlangen, Germany}
	\author{Roland Hayn}
	\affiliation{IM2NP, UMR CNRS 7334, Aix-Marseille Universit\'{e}, 13013 Marseille, France}

	\date{\today}
	
	\begin{abstract}
		The possibility of quantum anomalous Hall effect (QAHE) in two-dimensional kagome systems with $d$-orbital electrons is studied within a  multi-orbital tight-binding model. We concentrate on the case of isotropic Slater-Koster integrals which is realized in a recently discovered class of metal-organic frameworks TM$_3$C$_6$O$_6$ with transition metals (TM) in the beginning of the 3$d$ series. Furthermore, in the absence of exchange-type spin-orbit coupling, only isotropic Slater-Koster integrals give a perfect flatband in addition to the two dispersive bands hosting relativistic (Dirac) and quadratic band crossing points at high symmetry spots in the Brillouin zone. A quantized topological invariant requires a flux-creating spin-orbit coupling, giving Chern number (per spin sector) $C=1$ not only from the familiar Dirac points at the six corners of the Brillouin zone, but also from the quadratic band crossing point at the center $\Gamma$. In the case of isotropic Slater-Koster integrals the on-site spin-orbit coupling (SOC) is ineffective to create the QAHE and it is only the transfer or exchange-type SOC which can lead to a QAHE. Surprisingly, this QAHE comes from the nontrivial effective flux induced by the \textit{transverse} part of the spin-orbit coupling, exhibited by electrons in the $d$-orbital state with $m_l=0$ ($d_{z^2}$ orbital), in stark contrast to the more familiar form of QAHE due to the $d$-orbitals with $m_l \neq 0$, driven by the Ising part of spin-orbit coupling. The $C=1$ Chern plateau (per spin sector) due to Dirac point extends over a smaller region of Fermi energy than that due to quadratic band crossing. Our result hints at the promising potential of kagome $d$-electron systems as a platform for dissipationless electronics by virtue of its unique QAHE. 
	\end{abstract}
	
	\maketitle
	
	\section{\label{sec:level1}Introduction}
	
	Anomalous Hall effect (AHE) refers to a transport phenomenon where an electric current produces a voltage perpendicular to the former, in the absence of external magnetic field but with broken time-reversal symmetry. Usually occurring in ferromagnetic materials ~\cite{AHErmp}, its foundation was built in pioneering works ~\cite{KarplusLuttinger,Smit,Luttinger,Kondo,Berger,Nozieres}. More recent progress investigates this effect in magnetic semiconductors ~\cite{JungwirthPRL} and metallic ferromagnets~\cite{OnodaNagaosaJPSJ,KontaniPRB2007}. Theoretical studies of two-dimensional materials underline the significance of electronic structures with Dirac points and flatbands~\cite{Flatband1,Flatband2,Flatband3}, which have been reported in inorganic-metal compounds~\cite{DiracFlatbandKagome} as well as kagome metal-organic frameworks~\cite{Zhang,Hua,Shaiek} studied using \textit{ab-initio} first-principle calculations~\cite{Denawi}. These materials offer an alternative venue for such fascinating phenomena as high-temperature fractional quantum Hall effect ~\cite{Flatband3}, frustration-induced quantum spin liquid state~\cite{MakhfudzPRB2014,SavaryBalentsRPP}, magnetization plateaus from spin gap ~\cite{NatCommsHotta,CapponiPRB2013,MakhfudzPujolPRL2015}, and prospectively the AHE.  
	
	The AHE may arise from intrinsic mechanisms associated with non-collinear or chiral spin order ~\cite{Taguchi,YePRL,OnodaNagaosaPRL2,MartinBatista,OhgushiPRB,ZFangScience,CanalsPRB,MacDonaldPRL, BrunoPRL,MacDonaldPRL,MakhfudzPujolPRB2015} or from the Berry curvature in reciprocal $\mathbf{k}$ space ~\cite{SundaramNiu,Sinitsyn}, representing the topological property of the band structure of the electrons  ~\cite{Haldane2004PRL} characterized by Chern number~\cite{TKNN}, an integer~\cite{AHErmp}. The latter mechanism gives rise to the quantum anomalous Hall effect (QAHE) ~\cite{MacDonaldRMP} as shown theoretically in honeycomb ~\cite{Haldane1988PRL,OnodaNagaosaPRL1} and kagome~\cite{KSunPRL2009}lattices, magnetized graphene~\cite{QAHEgrapheneAFM, QAHEinGraphene}, semiconductor quantum wells ~\cite{CXLiuPRL,CXLiuAnnCondMat} and observed experimentally in magnetic topological insulators~\cite{Science,ScienceDeng}, so far relying on the gap at the Dirac or semi-Dirac point~\cite{Vanderbilt}. The QAHE 
	is also present in kagome systems either in organic~\cite{QAHE2DorganicTI,PRLabinitiMOF,SciRep} or inorganic frameworks~\cite{SCZhangPRL, Nature2018}, highlighting the importance of the effect in $d$-orbital systems (where the conduction electrons have angular momentum quantum numbers $l=2$ and $m_l=0,\pm 1,\pm 2$)~\cite{SCZhangPRL,Nature2018,QAHE2DorganicTI,Okamoto}, but an 
	analytical multi-orbital theory that probes the insulating Chern states at all nontrivial gaps induced by spin-orbit coupling (SOC) beyond the Ising approximation is still lacking. 
	
	This manuscript addresses the possible occurrence of quantum anomalous Hall effect in the $d$-orbital electron kagome systems, focusing on the Berry curvature intrinsic mechanism, based on an 
	analytical tight-binding model, in the isotropic limit of Slater-Koster integrals where a perfect flatband manifests. Such perfect flatband is ubiquitous in the electronic band structure of diverse families of compounds with kagome lattice structure.
	We will show that the isotropic limit of Slater-Koster integrals cannot give the QAHE effect with exclusively on-site SOC and we introduce also the transfer or exchange-type SOC into our model. 
	Going beyond the Ising limit of spin-polarized bands, we show that a quantum anomalous Hall state with $C=1$ per spin sector can arise from the Berry curvature of the bands of the  $m_l=0$ $d$-orbital ($d_{z^2}$ orbital) at their quadratic band crossing (QBC) point and, crucially, relies on the transverse part of the spin-orbit interaction that generates an orbital-dependent effective flux. This QAHE due to the QBC point accompanies the more familiar QAHE due to the Dirac point. For weak spin-orbit coupling, the band that was initially perfectly flat remains approximately flat in the presence of spin-orbit coupling, but now becomes topologically non-trivial as it carries non-zero Chern number. While our study relies on non-interacting electron model, the topologically non-trivial (nearly) flat band will also open up prospect for other exotic states beyond the Chern insulating state predicted in this work. 
	
	In Sections II and III, the model Hamiltonian and the resulting band structure and density of states are described. In Sections IV and V, Berry curvature and Chern number are defined, and the resulting QAHE is discussed. The following Sections VI and VII give the details on contributions of different $d$ orbitals to the QAHE and the effective flux mechanism that drives the QAHE. Section VIII exclusively discusses the role of onsite spin-orbit coupling. The paper ends with discussion and conclusions.
	
	\section{Model Hamiltonian}
\begin{figure}
		\includegraphics[angle=0,origin=c, scale=1.0]{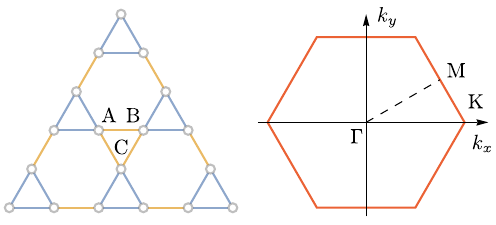}
		\caption{
			The kagome lattice (left) with its three sublattices and its first Brillouin zone (right) with the high-symmetry points.}
		\label{fig:KagomeLatticeandBZ}
	\end{figure}

	The conduction electrons of the kagome $d$-electron systems are described by a tight-binding Hamiltonian with nearest-neighbour hopping and spin-orbit coupling~:
	\begin{multline}
		\label{Hamiltonian}
		H=\sum_{\substack{\langle ij\rangle \\ \alpha\alpha'\sigma}}t_{ij,\alpha\alpha'}d^{\dag}_{i,\alpha\sigma}d_{j,\alpha'\sigma}+\mathrm{h.c.}
		+\sum_{\substack{i\\ \alpha\sigma}} E_{\alpha}d^{\dag}_{i,\alpha\sigma}d_{i,\alpha\sigma}
		\\
		-\sum_{\substack{i\\ \alpha\sigma\sigma'}} d^{\dag}_{i,\alpha\sigma}\left(\mathbf{M}\cdot\mathbf{s}\right)_{\sigma,\sigma'}d_{i,\alpha\sigma'}
		\\
		+\lambda_{OS}\sum_{\substack{i\\ \alpha\sigma \alpha'\sigma'}} d^{\dag}_{i,\alpha\sigma}\left(\mathbf{l}\cdot\mathbf{s}\right)_{\alpha\sigma,\alpha'\sigma'}d_{i,\alpha'\sigma'}
		\\
		+ i\lambda\sum_{\substack{\langle ij\rangle\\ \alpha\sigma,\alpha'\sigma'}} \nu_{ij}d^{\dag}_{i,\alpha\sigma}\left(\mathbf{l}\cdot\mathbf{s}\right)_{\alpha\sigma,\alpha'\sigma'}d_{j,\alpha'\sigma'}+\mathrm{h.c}\,\text{.}  
	\end{multline}
	$d^{\dag}_{i,\alpha\sigma}(d_{i,\alpha\sigma})$ are creation (annihilation) operators for $d$-electrons of spin $\sigma$ in orbital $\alpha$ and on site $i$ (Fig.\ref{fig:KagomeLatticeandBZ}), and the hopping matrix elements $t_{ij,\alpha\alpha'}$ in the first term are given by $t_{ij;\alpha,\alpha'}=f_{\sigma}V_{dd\sigma} + f_{\pi}V_{dd\pi} + f_{\delta}V_{dd\delta}$, where $f_{\sigma/\pi/\delta}$ are $\alpha,\alpha'$-dependent simple algebraic functions of nearest-neighbour vectors indices and the $V_{dd(\sigma/\pi/\delta)}$ are the Slater-Koster integrals for hopping of electrons between two $d$ orbitals that can be of $\sigma,\pi$, or $\delta$ type, available from the Slater-Koster table~\cite{Harrison,SupplementaryMaterials}. The second term accounts for the onsite energies which can adopt three different values: $E_1$ for the $d_{z^2}$-orbital, $E_2$ for the $d_{xz}$ and $d_{yz}$-orbital, and $E_{3}$ for $d_{xy}$ and $d_{x^2-y^2}$. The third term represents the spin splitting due to a ferromagnetic exchange field $\mathbf{M}$ arising from the time-reversal symmetry breaking ferromagnetic order of the spins of the transition metal ions, which we assume to be perpendicular to the lattice plane.  The last two contributions (onsite and transfer (exchange-type) SOCs) account for the intrinsic SOCs on a given site and between nearest neighbours. The latter term has to be preceded by an imaginary $i$ since the matrix elements $\nu_{ij}=\pm 1$ change sign according to whether the hopping occurs clockwise or counterclockwise around a triangle.   Note that the inclusion of the operator $\mathbf{l}$ allows us to extend the Kane-Mele SOC~\cite{KaneMelePRL} to multi-orbital systems and going past the Ising approximation of spin-polarized bands. 
	
The first term in Eq.(\ref{Hamiltonian}) simply describes kinetic hopping of electron between neighboring sites, which when alone would give rise to a metallic behavior. The onsite energy term represents the local energy of the electron that can be used to tune its energy level. The exchange-field term represents the effect of ferromagnetic ordering of localized ion spin on the electron energy, acting analogously to Zeeman coupling of external magnetic field to the electron spin. The last two terms reflect relativistic effect that couples the electron spin to its orbital angular momentum and gives rise to relativistic spin-splitting. The presence of the two types of spin-orbit coupling terms is mandated by intuitive physical consideration as follows. In a tight-binding model of electrons with spin and orbital degrees of freedom, electron hops between sites of the lattice. When the electron is at a site, it feels the electric field of the ion at the site, which manifests as relativistic onsite spin-orbit coupling, because the electric field transforms into an effective magnetic field in the electron's frame of reference, coupling to the electron spin, and the effective magnetic field is proportional to the orbital angular momentum of the electron. This is basically the standard atomic spin-orbit coupling of electron in isolated atom. On the other hand, the electron also hops between neighboring sites. During this hopping process, the electron feels the electric fields of the two ions located at the two neighboring sites. The electric fields transform into effective magnetic fields in the frame of reference of the hopping electron. The coupling between electron spin and the effective magnetic fields gives rise to the transfer (exchange type) spin-orbit coupling term. The latter is necessarily directional; it changes sign when reversing the direction of the electron hopping between the two sites.   
	
	\section{Band Structure and Density of States}
\begin{figure*}
		\includegraphics[angle=0,origin=c, scale=0.8]{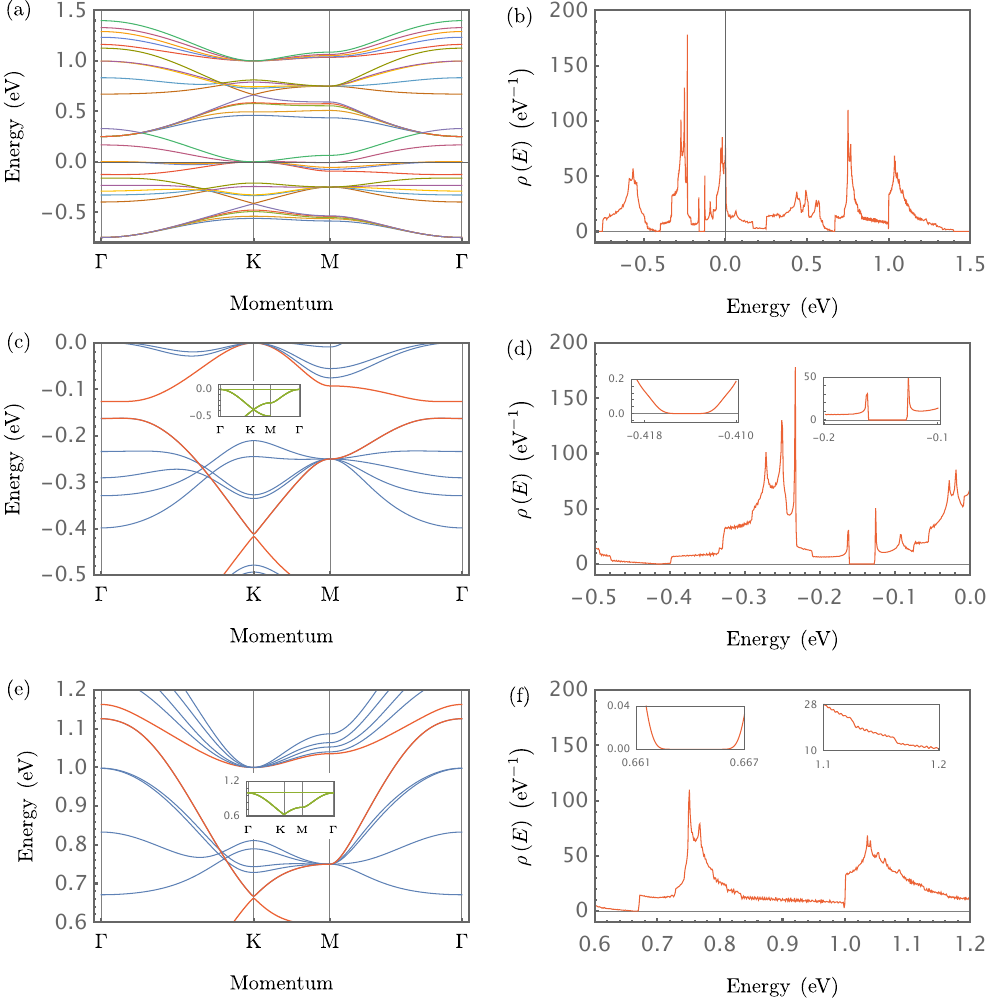}
		\caption{
			(a) Full band structure and (b) its corresponding density of states $\rho(E)$ from all five $d$-orbitals along the $K-\Gamma-M-\Gamma$ path in the first Brillouin zone (Fig.\ref{fig:KagomeLatticeandBZ}) for the kagome model Eq.(\ref{Hamiltonian}) with ~\cite{Note} $E_1=E_2=E_3=0.25$eV, $V_{dd\pi}=V_{dd\delta}=V_{dd\sigma}=-0.25$eV,   $\lambda=0.19$eV and $\lambda_{OS}=0.0$eV, and $M_z=1.0,M_x=M_y=0.0$ all in eV. (c) The low-energy part of the band structure corresponding to the spin-up sector, where the three bands of $d_{z^2}$ orbital are highlighted by red curves. The small inset shows the band structure without SOCs; i.e. $\lambda=\lambda_{OS}=0.0$eV, for which all the five $d$-orbitals become degenerate. (d) Corresponding density of states in the spin up sector. The visible intervals with vanishing density of states correspond to the gaps acquired by the ${d}_{z^2}$ orbital, at the Dirac and QBC points as marked by insets on the top left and right corners, respectively. The corresponding results for spin down sector are given in (e) and (f).}
		\label{fig:DensityofStates}
	\end{figure*}

	The physics of the conduction electrons can be deduced from the band structure computed from diagonalizing the Hamiltonian Eq.(\ref{Hamiltonian}) in the basis of the electron state (5 orbitals times 3 sublattices times 2 spin states). Presented in Fig.\ref{fig:DensityofStates} is the full band structure and its projections onto the spin up and spin down sectors (the two ``sectors'' are independent (decoupled), strictly speaking, only in the absence of spin-orbit coupling. But for simplicity, we continue to use the terminology even in the presence of spin-orbit coupling, justified when the spin-orbit coupling-induced gaps are much smaller than other energy scales). The spin up bands constitute the lower energy part of the full band structure, lower by typically few ($\sim 1.0$ to $3.0$) eV than the spin down state due to the exchange field $M_z$. At half filling, the spin down bands way up with high positive energy are not occupied, while the large gap $\sim M_z$ separating the spin up and down bands is trivial insulating gap. The band structure and density of states (and as discussed later, the Berry curvature, and Chern number) for the high energy spin down states are analogous to those of the spin up states, differing only by a shift in energy by $M_z$ corresponding to the trivial insulating gap separating them. 
	
	Our tight-binding calculations were focused on parameter regime which produces band structure with simple $s$-orbital electron-like profile, featuring a perfect flatband and two dispersive bands with two band crossing points; one with quadratic dispersion at $\Gamma$ and another with linear dispersion at K, in the energy range near zero energy (charge neutrality) especially in the beginning of 3$d$ series. These bands correspond to the spin up states of electron that would be occupied at zero doping corresponding to zero Fermi energy. For the quantum transport phenomenon of interest in this study, it is only these states that eventually participate in the transport; all other states far above zero energy corresponding to (unoccupied) spin down electron states and far below with electrons in spin up state do not participate in the transport and are not relevant to our study.
	
	The $s$-electron-like band structure can be produced by assigning equal values $V_{dd\sigma}=V_{dd\delta}=V_{dd\pi}=t_0$ for the Slater-Koster integrals in our tight-binding model, resulting in a perfectly flat band and two dispersive bands touching each other at the Dirac point at K and the QBC at $\Gamma$
	as shown in the insets of Fig. 2(c) and (e), to be compared with the bandstructures for Sc$_3$C$_6$O$_6$ and Ti$_3$C$_6$O$_6$ in Fig. 5 of Ref. ~\cite{Denawi}. 
	This isotropic limit, where $t_{ij,\alpha\alpha'}=t_0\delta_{\alpha\alpha'}$ is equal for all three nearest-neighbour vectors and also diagonal in orbital states $\alpha\alpha'$, will be the primary focus of this work, as it contains hidden extra spatial ($C_3$ point group) symmetry that distinguishes this ideal case from the anisotropic one, already considered for example in \cite{Okamoto}, where only kinetic hopping and onsite spin-orbit coupling terms were considered. As will be shown here later, the role of onsite spin-orbit coupling is muted in the isotropic limit. The on-site energies $E_{\alpha}$ separate the different $d$-orbitals from each other in the realistic band structure, but eventually,  since the onsite energies do not break time-reversal symmetry or spin-rotational symmetry, nor do they give rise to band dispersion or curvature, they do not directly influence the topological invariant of each band that characterizes the topological states of our interest.     
	
	Adding transfer SOC opens up a gap, as illustrated in Fig. 2(c)(e). The Ising part $l_z s_z$ of this SOC opens up a gap of order $\mathcal{O}(\lambda)$ at the gapless points, not just the Dirac point at K, but also the quadratic band crossing point at $\Gamma$, except for the $d_{z^2}$ orbital because of its $m_l=0$. For this ``trivial'' $d$-orbital, the planar part $l_xs_x+l_ys_y$ of the transfer SOC opens up a small gap at the quadratic band crossing at $\Gamma$ while opening a much smaller gap at the Dirac point at K~\cite{SupplementaryMaterials}. To probe the presence of a gap
	we compute the density of states
	with the result as shown in Fig. 2(b)(d)(f), where the gaps at the QBC and Dirac points give zero density of states when the gaps are not overshadowed by any bands. 
	
	Comparing the band structures for spin up and spin down sectors as shown in Fig. 2, it can be seen that while for spin up bands, the gaps at QBC and Dirac points are not overlapped by other bands, for spin down bands, only the gap at Dirac point remains free, while that at QBC is overshadowed by the top most band of $d_{z^2}$ orbital and several bands of the other $d$ orbitals, leading to finite density of states within the energy range of the QBC gap of the spin down sector. This situation persists at infinitesimally small spin-orbit coupling $\lambda$. This asymmetry between spin up and down bands evidently originates from the Zeeman-like exchange field $\mathbf{M}$ term, which breaks time-reversal symmetry. Turning off this term removes the asymmetry, giving rise to perfect Kramers degeneracy between the spin up and down bands.
	
	\section{Berry Curvature and Chern Number}
\begin{figure}
		\includegraphics[angle=0,origin=c, scale=1.15]{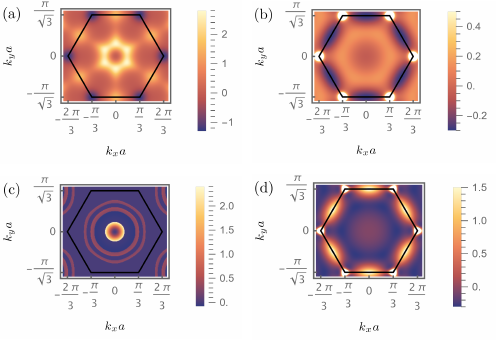}
		\caption{
			Color maps of the (dimensionless) Berry curvature field $a^{-2}\Omega(k_xa, k_ya)$ where $a$ is lattice spacing,  taken as the reference unit of length, with the same parameters as those for Fig. 2 evaluated at (a)$E_F=-0.150$eV for the QBC and (b)$E_F=-0.414$eV for the Dirac point, within the gaps of the band structure for spin up sector with SOC as shown in Fig. 2. The corresponding results for spin down bands are given in (c) and (d) at energies $E_F=1.16$eV and $E_F=0.665$eV, respectively. The bright areas of $a^{-2}\Omega(k_xa, k_ya)$ are located (a)(c) around the center of BZ ($\Gamma$ point) and (b)(d) at the corners of BZ and are responsible for the Chern plateaus in Fig.\ref{fig:ChernNumber}.}
		\label{fig:BerryCurvatureField}
	\end{figure}

	In the intrinsic mechanism for the anomalous Hall effect, the transverse motion of the electron is caused by the ``anomalous velocity'', proportional to the cross product of the electric field and the Berry curvature $\Omega(\mathbf{k})$ defined in momentum space ~\cite{Haldane2004PRL}. This Berry curvature acts as an effective magnetic field in momentum space. Integrating this field over a surface in 2D $\mathbf{k}$ space gives rise to a topological magnetic charge, which is essentially the Chern number characterizing the anomalous quantum Hall effect ~\cite{TKNN}. At arbitrary temperature $T$, the Berry curvature field is given by ~\cite{AHErmp}
	\begin{equation}\label{InterbandChernGeneralT}
		\Omega(\mathbf{k})=\hbar^2\sum_{n\neq n'}G_{nn'}(\mathbf{k}) \mathrm{Im}\left[\frac{\langle\psi_{n\mathbf{k}}|v_x(\mathbf{k})|\psi_{n'\mathbf{k}}\rangle \langle\psi_{n'\mathbf{k}}|v_y(\mathbf{k})|\psi_{n\mathbf{k}}\rangle}{\left(\varepsilon_{n'\mathbf{k}}-\varepsilon_{n\mathbf{k}}\right)^2}\right]
	\end{equation}
	with
	\begin{equation}
		G_{nn'}(\mathbf{k})=f(\epsilon_n(\mathbf{k}))-f(\epsilon_{n'}(\mathbf{k}))
	\end{equation}
	where $f(\epsilon_n(\mathbf{k}))=(\exp(\beta(\epsilon_n(\mathbf{k})-E_F))+1)^{-1}$ for which $\beta=1/(k_BT)$ with Boltzmann constant $k_B$ and reduced Planck constant $\hbar$, $\mathrm{Im}$ indicates the imaginary part of its argument, $\psi_{n\mathbf{k}},\varepsilon_{n\mathbf{k}}$ are respectively the eigenfunction and eigenvalue of the $n^{\mathrm{th}}$-eigenstate of $H_{\mathbf{k}}$ while the velocity operator is given by $v_x=(1/\hbar)\partial H_{\mathbf{k}}/\partial k_x$ and similarly for the $y$ part. The profile of the Berry curvature field is illustrated in Fig.\ref{fig:BerryCurvatureField}. 
	
	The corresponding  Chern number is then evaluated over the Brillouin zone
	\begin{equation}\label{InterbandChernT=0}
		C=\frac{1}{2\pi}\int_{\mathrm{BZ}} d^2\mathbf{k}\Omega(\mathbf{k})=\frac{\Phi}{2\pi}
	\end{equation}
	linked to the Berry phase $\Phi$ ~\cite{Haldane2004PRL}, giving an intrinsic Hall conductivity $\sigma_{xy}=(e^2/h)C$ where $e$ is the elementary charge of electron and $h$ is the Planck constant. As shown in ~\cite{AHErmp}, the Chern number can also be recast in terms of intraband processes and computed efficiently using a trick based on a unitary transformation function ~\cite{Fukui}. In the present work, the interband definition of Chern number is used.   
	
	\begin{figure*}
		\includegraphics[angle=0,origin=c, scale=1.0]{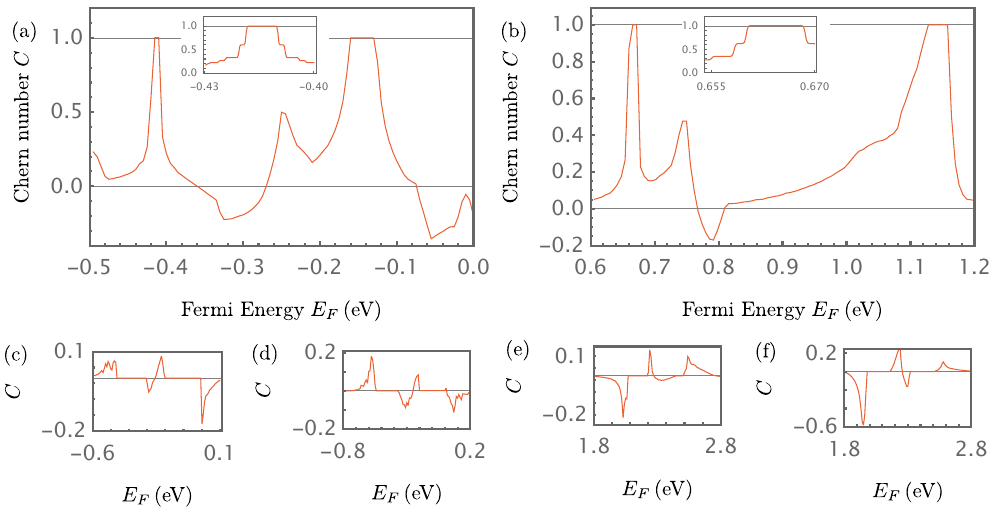}
		\caption{
			(a) Dependence of the Chern number $C$ on the Fermi energy $E_F$ (eV) at $T=0.0001$eV($=1.1$Kelvin) in the spin up sector for kagome metal-organic framework with the same parameters as those for Figs. 2-3. Clear plateaus at $C=1$ fall within the gaps of the $d_{z^2}$ orbital, centered at about -0.150eV for gapped QBC point at $\Gamma$ and -0.414eV for gapped Dirac point at K in Fig.\ref{fig:DensityofStates}. As comparison, the results with parameters that highlight the contributions of (c)$d_{xz},d_{yz}$ and (d)$d_{xy},d_{x^2-y^2}$ orbitals are presented, both of which give trivial $C=0$ insulating states, as will be described in details in Section VI. The corresponding results for spin down sector are shown in (b)(e)(f).}
		\label{fig:ChernNumber}
	\end{figure*}
	
	\section{Overall Result on Predicted Quantum Anomalous Hall Effect and Its Physical Origin}
	The  Chern number is computed from the full $30\times 30$ Hamiltonian matrix using Eq.(\ref{InterbandChernGeneralT}), focusing on the spin up sector just below the zero Fermi energy, at $T=0.0001$eV ($=1.1$Kelvin), low enough to enter the quantum regime while still being within reach of existing cryogenic technology.
	
	The resulting profile of the Chern number is shown in Fig. \ref{fig:ChernNumber} as function of the Fermi energy. The computed topological invariant (proportional to the anomalous Hall conductivity) is mostly nonzero and varies continuously with Fermi energy, but may display exact quantization (integer Chern number) within the spin-orbit coupling-induced gap, appearing as plateau. 
	These plateaus are clearly visible in Fig. \ref{fig:ChernNumber}, showing Chern number at quantized value $C=1$, corresponding to a quantized Hall conductivity $\sigma_{xy}=e^2/h$. It can be verified by comparing with Fig. 2(c)(e) that the plateaus fall within the gaps of the band structure of the $d_{z^2}$ orbital in the corresponding spin sector. 
	
	Interestingly, one of the  $C=1$ Chern plateaus comes from the QBC point 
	at the center $\Gamma$ of the Brillouin zone, as reflected by the peak of the Berry curvature field at the center in Fig. 3(a)(c), and thus corresponds to non-relativistic electrons associated with the normally considered irrelevant $d_{z^2}$ orbital. This $C=1$ Chern plateau corresponds to a Berry phase $\Phi=2\pi$ originating from a QBC point in a system with $C_6$ symmetry~\cite{KSunPRL2009}. The associated QAHE is robust, giving a Chern plateau whose width is proportional to $\lambda$. Noting that the Chern plateau occurs just (slightly) below the zero Fermi energy, this implies that the QAHE requires hole doping concentration corresponding to 1/3 filling factor into the system, because one has to deplete the flatband of electrons while leaving the lower two bands occupied (see the inset of Fig. 2(c) for example).
	
	The gapped Dirac points corresponding to the $d_{z^2}$ orbital, located at the six corners ($\pm$K) of the Brillouin zone, also display a $C=1$ Chern plateau. Due to the smaller gap at the Dirac point, the plateau lies in a tiny region of Fermi energy with much smaller width (by an order of magnitude or more) than that at the QBC. More precisely, for reasonable sets of parameters  with $\lambda$ a few hundreds meV as used to produce the figures in this article, the gap at the Dirac point is less than 10 meV, whereas that at the QBC is several tens of meV, as illustrated in Fig. \ref{fig:ChernNumber}~\cite{SupplementaryMaterials}. In real materials, SOC is usually weak but the Chern plateau at the QBC gap is still expected to be broader than the one at the Dirac gap, thus rendering the QAHE from the QBC point more prominent. Furthermore, since the energy of the Dirac point is further below the 
	Fermi energy (see Fig. 2(c), corresponding to $1/3$ filling factor of the spin up band, the QAHE from such Dirac point would require an amount of hole doping equivalent to 2/3 filling factor (because one has to vacate both the flat band and middle band of electrons while leaving the lowest band occupied (inset of Fig. 2(c)), larger than that for QAHE due to QBC. Hence, in materials with weak SOC, the QBC Chern insulating state will be more dominant than that due to the Dirac point.  
	
	\begin{figure}
		\includegraphics[angle=0,origin=c, scale=1.0]{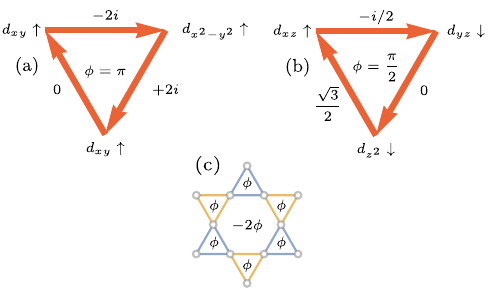}
		\caption{
			Representative effective fluxes induced by the (a) Ising part and (b) transverse part of exchange SOC, with the orbital state at the site and the resulting (generally complex) number indicating the overlap $\langle \alpha,s|l_as_a|\alpha',s'\rangle$ where $\alpha,s$ are the orbital, spin indices (marked by spin up or down arrow) while $a=x,y,z$ and (c) the resulting flux pattern. Only when the hopping is induced by the transverse part of the exchange SOC $l_xs_x+l_ys_y$ involving spin flip and a $d_{z^2}$ orbital state, will $\phi$ be nontrivial $\phi\neq 0,\pi$ and thus breaks time-reversal symmetry, as exemplified in (b). All other contributions give trivial (non-time-reversal symmetry-breaking) fluxes, an exact result in the isotropic Slater-Koster integrals limit, as described in more details in Sections VII and VIII.}
		\label{fig:FluxPattern}
	\end{figure}
	
	On the other hand, the supposedly non-trivial $d$-orbitals: $d_{xz}$, $d_{yz}$, $d_{xy}$, and $d_{x^2-y^2}$ orbitals are found to give trivial  Chern ($C=0$) plateau within the gaps, as illustrated in Fig. \ref{fig:ChernNumber}(c)-(f), at \textit{any} spin-orbit coupling strength $\lambda$; this clearly relates to a non-perturbative
	topological origin and in fact pertains to the effective flux, as described in more details in Section VII. More precisely, adopting the flux analysis method~\cite{KontaniJPSJ762007main,KontaniPRL2008main}, for the kagome lattice the electron hopping induces fluxes $\phi$ through the triangles and $-2\phi$ through the hexagons,  giving zero average flux per unit cell
	~\cite{OhgushiPRB}. The phase $\phi$ depends on the combination of the orbitals, resulting in time-reversal symmetry breaking, unless $\phi$ is 0 or $\pi$ (referred here as ``trivial''). The fluxes turn out to be trivial for  exchange-type SOC-induced hoppings involving only the ``nontrivial'' $d$-orbitals ($d_{xz}, d_{yz}, d_{xy}, d_{x^2-y^2}$) whereas the fluxes are nontrivial for spin-flipping hoppings between the $d_{z^2}$ orbital and the other four orbitals, made possible by the transverse part of the SOC $l_xs_x+l_ys_y$, as illustrated in Fig.\ref{fig:FluxPattern}. This explains why the $d_{z^2}$ orbital gives rise to the QAHE with its Chern plateaus. Due to the topological flux origin of the Chern insulating state where the flux is independent of the strength of SOC, the state exists at arbitrarily small but nonzero $\lambda>0$, with $C=1$ Chern plateaus at the QBC and Dirac points of the $d_{z^2}$ orbital.
	
	As illustrated in Fig. 4(b)(e)(f) for the spin down bands, the Chern number vs. Fermi energy qualitatively has the same profile as that of the spin up bands. In particular, there are two $C=1$ peaks corresponding to the Dirac and QBC gaps of the $d_{z^2}$ orbital, while all the other $d$ orbitals give trivial $C=0$ Chern plateaus. The robustness of the $C=1$ peak at QBC of the spin down band of the $d_{z^2}$ is especially remarkable in view of the fact that for spin down state, as shown in Fig. 2(e), the QBC gap of the $d_{z^2}$ orbital is overshadowed by the top most band of the $d_{z^2}$ orbital itself (marked by red curve) and those of the other $d$ orbitals. Crucially, the top most band of $d_{z^2}$ orbital does not destroy the $C=1$ peak of the QBC gap because the Berry curvature of this band is concentrated at the $\Gamma$ point (Fig. 3(c)), which corresponds to energy very close to the upper edge of the QBC gap. On the other hand, since all the other $d$ orbitals carry trivial (i.e. zero) Chern number, they also do not affect the topological character of the $d_{z^2}$, leaving the $C=1$ peak at the QBC gap of this latter orbital intact. This observation simply demonstrates the robustness of the predicted Chern insulating state from the $d_{z^2}$ orbital.   
	
	For three candidate materials in the  metal-organic frameworks TM$_3$C$_6$O$_6$ family with TM = Sc, Ti, V, \textit{ab-initio} band structure calculation ~\cite{Denawi} indicates that all of them have some finite densities of states for the $d_{z^2}$ orbital at or just below the zero Fermi energy, with  Sc$_3$C$_6$O$_6$ having the most dominant content of $d_{z^2}$. This suggests that the predicted QAHE based on $d_{z^2}$ orbital is expected to be applicable to these real kagome organic metal compounds.   
	
	More detailed analysis of the contributions of different $d$ orbitals to the QAHE and the flux mechanism that induces this effect are given in the following sections.
	
	\section{Contribution of Different $d$-Orbitals to Quantum Anomalous Hall Effect}
	In this section, we elucidate the contribution of different $d$-orbitals to quantum anomalous Hall effect. Different $d$-orbitals contribute differently to the quantum anomalous Hall effect; some contribute significantly, the other do not contribute at all. To show this, we compute the Chern number that characterizes the quantum anomalous Hall effect by choosing the parameters appropriately in such a way that the contribution of each $d$-orbital can be isolated. For the sake of conciseness however, we assume degeneracy between symmetry-related orbitals; the $d_{xz}$ orbital is assumed to be degenerate with the $d_{yz}$ orbital while the $d_{xy}$ orbital is assumed to be degenerate with the $d_{x^2-y^2}$ orbital. When we identify the contribution of an orbital $d_{\alpha}$, we set the onsite energy of this orbital (and its degenerate partner) to be significantly lower than the remaining orbitals and we compute the Chern number in the Fermi energy range within the energy range of the orbital $d_{\alpha}$ under consideration. 
	
\begin{figure}
		\includegraphics[angle=0,origin=c, scale=0.7]{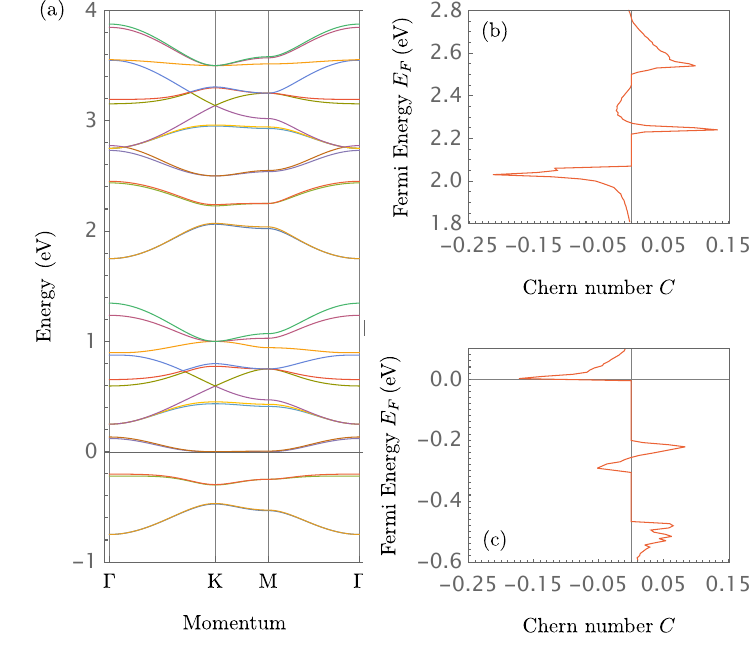}
		\caption{
			(a) Band structure (lower half) in the energy range corresponding to the $d_{xz}$ and $d_{yz}$ orbitals  and (c) the contribution of the $d_{xz}$ and $d_{yz}$ orbitals to the Chern number $C$ as a function of the Fermi energy $E_F$ (eV) at $T=0.0001$eV($\equiv 1.1$Kelvin) for the tight-binding model Eq.( \ref{Hamiltonian}) of the kagome metal-organic-framework, in the spin up sector. To separate out these two $m_l=+/-1$ orbitals, we have set $E_1=E_3=2.0$eV, $E_2=1.0$eV, while using $V_{dd\pi}=V_{dd\delta}=V_{dd\sigma}=-0.25$eV, $ \lambda_{OS}=0.0$eV, $M_x=M_y=0.0,M_z=2.5$eV with $\lambda=0.2$eV, where all five orbitals are included. The $C=0$ plateaus fall within the gaps of the $d_{xz}$ and $d_{yz}$ orbitals' band structure. The corresponding profiles for the band structure of the spin down sector and its Chern number are respectively given in the upper half of (a) and (b).}\label{fig:ChernNumberXZYZ}
	\end{figure}

	The result of the calculation of Chern number $C$ vs Fermi energy $E_F$ from $d_{xz}$ and $d_{yz}$ orbitals is shown in Fig. \ref{fig:ChernNumberXZYZ}. As can be seen, these orbitals do not display Chern plateaus even with rather strong spin-orbit coupling $\lambda=0.5$eV. In fact, the Chern peaks that they display as function of Fermi energy never exceed unity in absolute value, which cannot qualify them as nontrivial Chern number. Therefore, it can be concluded that these two orbitals do not contribute to the quantum anomalous Hall state.
	
\begin{figure}
		\includegraphics[angle=0,origin=c, scale=0.7]{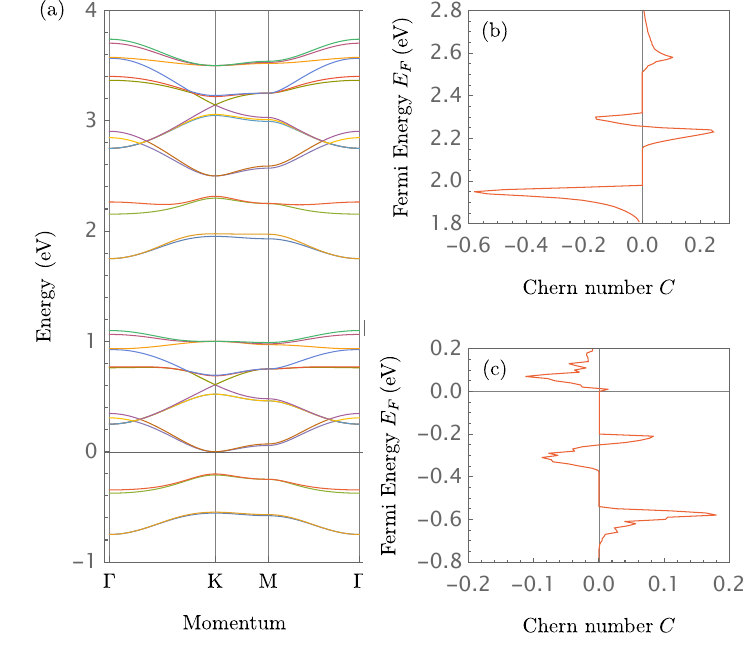}
		\caption{
			(a) Band structure (lower half) and (c) contributions of the  $d_{xy}$ and $d_{x^2-y^2}$ orbitals to the Chern number $C$ as a function of the Fermi energy $E_F$ (eV) at $T=0.0001$eV($\equiv 1.1$Kelvin) for the tight-binding model Eq.( \ref{Hamiltonian}) of the kagome metal-organic-framework, for spin up sector. To separate out these $m_l=+/-2$ orbitals, we have set $E_1=E_2=2.0$eV, $E_3=1.0$eV, while using $V_{dd\pi}=V_{dd\delta}=V_{dd\sigma}=-0.25$eV, $ \lambda_{OS}=0.0$eV, $M_x=M_y=0,M_z=2.5$eV with $\lambda=0.2$eV  where all five orbitals are included. Chern plateaus with $C=0$ appear within the band gaps: the one in the interval $(-0.2,0.0)$eV is from gapped Dirac point while the other in the interval $(-0.6,-0.4)$eV is from gapped quadratic band crossing point. The corresponding results in the spin down sector are given respectively in the upper half of (a) for the band structure and (b) for the Chern number.}\label{fig:BandStructure+ChernNumberXYX2Y2moderateSOC}
	\end{figure}

	The contribution of the $d_{xy}$ and $d_{x^2-y^2}$ orbitals on the Chern number vs Fermi energy is shown in Fig. \ref{fig:BandStructure+ChernNumberXYX2Y2moderateSOC}. At realistic small values of spin-orbit coupling $\lambda\lesssim 0.3$eV, plateaus are visible, but with Chern number $C=0$. This suggests that these two orbitals have no contribution to quantum anomalous Hall effect in our system. As shown in Fig. \ref{fig:BandStructure+ChernNumberXYX2Y2moderateSOC} and discussed further in the next section, even a rather strong spin-orbit coupling $\lambda\gtrsim 0.5$eV gives plateaus with zero Chern number at Dirac point. The  $d_{xy}$ and $d_{x^2-y^2}$ orbitals therefore do not give quantum anomalous Hall effect in our system.
	
	Finally, the Chern number contribution from $d_{z^2}$ orbital basically takes profile as shown in Fig.~\ref{fig:ChernNumber}(a)-(b). Two plateau peaks appear; one very sharp peak (i.e. very narrow plateau) at Dirac point and a broader plateau at QBC point.
	
	\section{Physical Origin of the Quantum Anomalous Hall Effect: The Effective Flux  due to the Exchange-Type Spin-Orbit Coupling}
	In the previous sections, spin-orbit coupling is shown to have strong influence on the band structure, the precise nature of which depends on each particular $d$ orbitals. In addition, the spin orbit-coupling also has strong influence on the Chern number contribution of each orbital, and therefore, its role in quantum anomalous Hall effect.  We have also confirmed numerically that the absence of spin-orbit coupling (setting $\lambda=0$) gives zero Chern number, in agreement with expectation that nonzero Chern number requires a gap within which the Fermi energy should fall. 
	
	As stated at the end of the previous section, for $d_{z^2}$ orbital, Chern plateau with $C=1$ appears in the presence of exchange-type spin-orbit coupling, at the gap associated with the QBC point at $\Gamma$ in the absence of spin-orbit coupling. As stressed earlier in Section V, this $C=1$ QAHE can be interpreted as arising from the $\Phi=2\pi$ Berry phase of the QBC point, thus guaranteeing the stability and robustness of the $C=1$ Chern insulating state from the QBC point. The gapped Dirac point of the bands of the $d_{z^2}$ orbital also displays $C=1$ Chern plateau but with much narrower band width, due to its much smaller gap. 
	
	On the other hand, the Chern number of Chern plateaus due to $d_{xy}$ and $d_{x^2-y^2}$ orbitals only give $C=0$ at any value of spin-orbit coupling, both at the gapped Dirac point and quadratic band crossing point, as illustrated in Fig.\ref{fig:BandStructure+ChernNumberXYX2Y2moderateSOC}. This demonstrates a completely different scenario for the realization in QAHE compared to those known in existing literature, where QAHE normally comes from the inter-orbital transition between $m_l=\pm 1$ $d$-orbitals ($d_{xz}$ and $d_{yz}$ orbitals) or between $m_l=\pm 2$ $d$-orbitals ($d_{xy}$ and $d_{x^2-y^2}$ orbitals)\cite{KontaniPRB2007}\cite{SCZhangPRL}. This novel scenario is made possible by the geometry of the kagome lattice which involves triangles, which renders the hopping between nearest-neighbour sites to have both $x$ and $y$ components in general, allowing for contribution to transverse charge current even from $m_l=0$ $d$-orbital ($d_{z^2}$ orbital). 
	
\begin{figure}
			\includegraphics[angle=0,origin=c, scale=1.0]{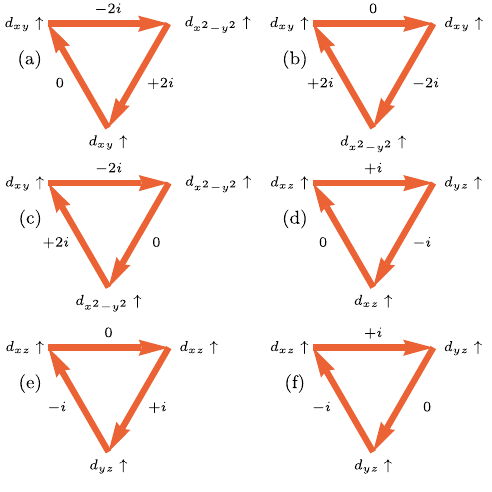}
			\caption{
				(a)-(f) Several possible flux patterns in a ``down triangle'' of the kagome lattice due to the clockwise hopping of electrons with orbitals $d_{\alpha}(d_{\alpha'})$ at site $i(j)$ respectively  in the low energy (occupied) spin up sector, induced by the Ising part $l_zs_z$ of the exchange-type spin-orbit coupling (no spin flip during the hopping). The generally complex number $z_{ij}$ attached to each arrow pointing from site $i$ to site $j$ comes from the overlap factor $\langle d_{\alpha'}|l_z| d_{\alpha'}\rangle$ evaluated across the nearest-neighbour bond $\langle ij\rangle$. The arrow with 0 implies that the hopping is induced by the kinetic term rather than the spin-orbit coupling term. The flux $\phi$ is given by $\phi=\sum_{\langle ij\rangle}(\Tilde{\phi}_{ij}+\pi/2)$ where the sum is around the loop traversing the triangle clockwise and the complex factor at each arrow is written as $z_{ij}=|z_{ij}|\exp(i\Tilde{\phi}_{ij})$ with $\phi_{ij}=\Tilde{\phi}_{ij}+\pi/2$ as given in Eq.(\ref{PhaseSOC}) with $a=z$ and $\nu_{ij}=1$. The net flux through the triangle is zero in all these patterns. This result is exact in the isotropic limit of Slater-Koster integrals $V_{dd\sigma}=V_{dd\delta}=V_{dd\pi}$.}\label{fig:FluxPatternExchangeSOC}
		\end{figure}

	A theoretical explanation is provided as to why the gaps associated with the ``non-trivial $d$-orbitals'' do not contribute to the QAHE, giving $|C|=1$ as occurring in kagome flux model \cite{OhgushiPRB}. The key answer is that the flux pattern induced by the exchange-type spin-orbit coupling must satisfy the condition of zero net flux or zero average flux, as pointed out in Haldane model \cite{Haldane1988PRL} and satisfied also in kagome flux model \cite{OhgushiPRB}. Unfortunately, this condition is not satisfied by the non-trivial orbitals in our model, as described in details in the following.
	Consider a triangle in a kagome lattice. The main contribution to the flux is provided by hopping between nearest-neighbour sites induced by the exchange-type spin-orbit coupling, combined with the complex phase factor provided by the overlap between $d$-orbitals of similar types; $d_{xz}-d_{yz}$ orbitals, and $d_{xy}-d_{x^2-y^2}$ orbitals. These pairs of $d$-orbitals give rise to the following complex factor for spin up electrons;
	
	\begin{equation}\label{overlapfirst}
		\langle d_{xz},s_z=+\frac{1}{2}|l_z|d_{yz},s_z=+\frac{1}{2}\rangle =-i,
	\end{equation}
	\begin{equation}
		\langle d_{yz},s_z=+\frac{1}{2}|l_z|d_{xz},s_z=+\frac{1}{2}\rangle =+i,
	\end{equation}
	\begin{equation}
		\langle d_{xy},s_z=+\frac{1}{2}|l_z|d_{x^2-y^2},s_z=+\frac{1}{2}\rangle =2i,
	\end{equation}
	\begin{equation}\label{overlaplast}
		\langle d_{x^2-y^2},s_z=+\frac{1}{2}|l_z|d_{xy},s_z=+\frac{1}{2}\rangle =-2i,
	\end{equation}
	where the bra and ket states are assumed to have the spin up state (the spin down state will give additional minus sign to the right hand side of the equations).
	Several possible flux patterns induced by exchange-type spin-orbit coupling around a ``down triangle'' from clockwise hopping of $d$-orbital electrons in the kagome lattice within our model are illustrated in Fig.\ref{fig:FluxPatternExchangeSOC}. The flux factor associated with an arrow that goes from site $i$ to $j$ due to the exchange-type spin-orbit coupling arises from the overlap between the wave functions of the electrons from two $d$-orbital states; between the orbital $d_{\alpha}$ at site $i$ and the orbital $d_{\alpha'}$ at site $j$. It can be seen that the net flux through the triangle is zero in all the patterns shown in Fig.\ref{fig:FluxPatternExchangeSOC}. 
	
	One can draw similar figures of flux pattern for the counter clockwise hopping of electron or the flux pattern for the ``up triangle''. Finally, one can deduce the flux pattern for the hexagon, assuming a flux pattern where all down triangles have identical flux pattern among themselves and all up triangles have identical flux pattern among themselves. One will find that the flux through the hexagons is zero. One can also consider the overlap of these orbitals over the transverse part of the spin-orbit coupling; 
	$\langle d_{\alpha}, s_z|l_xs_x+l_ys_y|d_{\alpha'},s'_z\rangle$ where $\alpha,\alpha'=xz,yz$ or $xy,x^2-y^2$ and $s_z,s'_z=\pm 1/2$. It is easy to see that the overlap is always zero.
	
	It can therefore be concluded that the exchange-type spin-coupling gives zero flux at triangles and hexagons when only these ``non-trivial'' $d$-orbitals are involved. Despite having zero net flux, such trivial flux state does not have the flux structure needed to host a quantum Hall effect like that in the kagome flux model\cite{OhgushiPRB}. In fact, according to Haldane model, zero flux ($\phi=0$) gives $C=0$ from the gapped Dirac points. In other words, zero flux renders the paired Dirac points at $\pm K$ to always have gaps of the same sign. The exchange-type spin-orbit coupling in our model thus does not give rise to a QAHE from the Dirac points of the nonzero $m_l=\pm 1,\pm 2$ ($d_{xz}, d_{yz}, d_{xy}, d_{x^2-y^2}$) $d$-orbitals in the band structure. This analysis explains trivial Chern plateaus ($C=0$) observed in the figures of previous section coming from the nontrivial $d$-orbitals.  
	
\begin{figure}
			\includegraphics[angle=0,origin=c, scale=1.0]{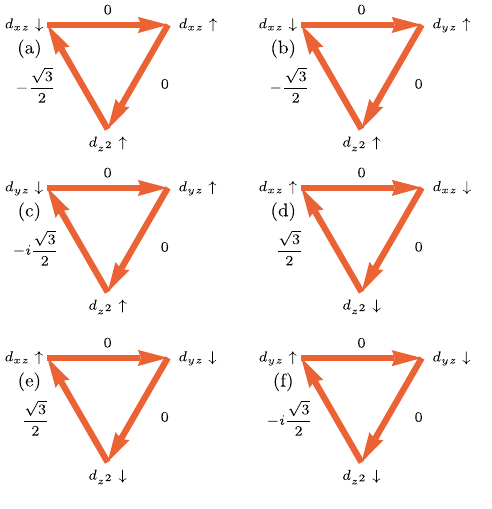}
			\caption{
				(a)-(f)Several possible flux patterns in a ``down triangle'' of the kagome lattice due to the clockwise hopping of electrons with orbitals $d_{\alpha}(d_{\alpha'})$ at site $i(j)$ respectively, induced by the transverse part $l_xs_x+l_ys_y$ of the exchange-type spin-orbit coupling (involving spin flip during the hopping).} \label{fig:FluxPatternExchangeTransverseSOCwithZ2orbital}
		\end{figure}

	The non-vanishing contribution of the $d_{z^2}$ orbital to the quantum anomalous Hall effect (and its dominance, since all the other orbitals only give trivial $C=0$ plateaus) is made possible by the transverse part of the spin-orbit coupling $l_x s_x+l_y s_y$, which is completely missed by the existing works in the literature, which normally consider only the Ising part of the spin-orbit coupling $l_z s_z$ and which therefore omit the $d_{z^2}$ orbital because of its $m_l=0$. In fact, the $C=1$ Chern-insulating state can be explained by the same flux argument, but this time involving the overlaps of the $d_{z^2}$ orbital with all the other orbitals over the $l_x s_x+l_y s_y$ operator as given in the following
	
	\begin{equation}\label{z2overlapfirst}
		\langle d_{z^2},s_z=+\frac{1}{2}|l_x s_x+l_y s_y|d_{xz},s_z=-\frac{1}{2}\rangle =-\frac{\sqrt{3}}{2},
	\end{equation}
	\begin{equation}
		\langle d_{z^2},s_z=+\frac{1}{2}|l_x s_x+l_y s_y|d_{yz},s_z=-\frac{1}{2}\rangle =i\frac{\sqrt{3}}{2},
	\end{equation}
	\begin{equation}
		\langle d_{z^2},s_z=+\frac{1}{2}|l_x s_x+l_y s_y|d_{xy},s_z=-\frac{1}{2}\rangle =0,
	\end{equation}
	\begin{equation}\label{z2overlaplast}
		\langle d_{z^2},s_z=+\frac{1}{2}|l_x s_x+l_y s_y|d_{x^2-y^2},s_z=-\frac{1}{2}\rangle =0,
	\end{equation}
	while noting that
	\begin{equation}\label{lastparttransverse}
		\langle d_{yz},s_z=+\frac{1}{2}|l_x s_x+l_y s_y|d_{xz},s_z=-\frac{1}{2}\rangle =0.
	\end{equation}
	It is already clear from the flux factor involving $d_{z^2}$ orbital in Eqs.(\ref{z2overlapfirst}-\ref{z2overlaplast}) that considering a triangle where an electron hops between the sites of the triangle (while flipping its spin) in such a way that it takes $d_{z^2}$ orbital state at one of the sites and $d_{xz}$ or $d_{yz}$ orbital states at the two other sites will give a nonzero net flux (that is, a phase factor $\exp(i\phi)$ to the electron wave function, where $\phi$ is the net flux through the triangle). The corresponding flux in the hexagon is $-2\phi$ so that the net (or average flux) is zero in the unit cell. This flux is the origin of the $C=1$ Chern insulating state due to the $d_{z^2}$ orbital, just like that in Haldane model \cite{Haldane1988PRL} and kagome flux model \cite{OhgushiPRB}. Examples of triangle with nonzero flux pattern are shown in Fig. \ref{fig:FluxPatternExchangeTransverseSOCwithZ2orbital}. 
	
		To characterize the flux induced by spin-orbit coupling, two alternative definitions of the flux phase are introduced. First, the flux phase is defined with respect to the total effective hopping induced by the kinetic hopping and spin-orbit coupling. Define the total effective hopping integral as follows,
		\begin{equation}
			\Tilde{t}_{ij}=t_{ij}+\delta t_{ij}=|\Tilde{t}_{ij}|e^{i\psi_{ij}},
		\end{equation}
		where $t_{ij}$ is due to the kinetic hopping term in Eq.( \ref{Hamiltonian}) and is real valued by definition, whereas $\delta t_{ij}$ is due to the spin-orbit coupling terms. In general, the total effective hopping integral $\Tilde{t}_{ij}$ is a complex quantity due to the complex valued-ness of the $\delta t_{ij}$,
		\begin{equation}
			\delta t_{ij}=|\delta t_{ij}|e^{i\phi_{ij}}
		\end{equation}
		permitting us to define the complex phase $\psi_{ij}$ coming from hopping from site $i$ to site $j$. The resulting flux through a triangle in the kagome lattice is then given by 
		\begin{equation}\label{phasepsi}
			\psi=\sum_{ij\in\triangle,\triangledown}\psi_{ij}  \; , \; \; \;\;\tan(\psi_{ij})=\frac{|\delta t_{ij}|\sin\phi_{ij}}{t_{ij}+|\delta t_{ij}|\cos\phi_{ij}}.
		\end{equation}
		The second definition of the flux phase is simply derived from the phase angle of the hopping integral contribution from the spin-orbit coupling terms,
		\begin{equation}\label{phasephi}
			\phi=\sum_{ij\in\triangle,\triangledown}\phi_{ij}
		\end{equation}
		which depends on whether the spin-orbit coupling is of transfer (exchange) type or onsite type. As can be seen from Eqs.(\ref{phasepsi}) and (\ref{phasephi}), the two definitions are related to each other.
		
		In this section,  we evaluate the two flux phase definitions involving transfer (exchange) type spin-orbit coupling.
		In this case, the phase induced by the spin-orbit coupling in the hopping of the electron from site $i$ to site $j$ is given by
		\begin{equation}\label{PhaseSOC}
			\exp(i\phi_{ij})=\frac{_j\langle \alpha,s|il_as_a\nu_{ij}|\alpha',s'\rangle_i}{|_j\langle \alpha,s|il_as_a\nu_{ij}|\alpha',s'\rangle_i|}
		\end{equation}
		where the subscripts $i,j$ represent site index while the $i$ in $il_as_a$ is the unit imaginary number whereas $a=x,y,z$ represents the component of the spin-orbit coupling. Let us first focus on the Ising part, i.e. $a = z$. It is noted from Eq.(\ref{overlapfirst}) that $_j\langle \alpha,s|l_z|\alpha',s'\rangle_i$ is purely imaginary or zero (for $d_{z^2}$ orbital, where $m_l=0$). As such, since $\nu_{ij}=\pm 1$, then $_j\langle \alpha,s|il_as_a\nu_{ij}|\alpha',s'\rangle_i$ is purely real and so $\phi_{ij}=0,\pi$, corresponding respectively to the positive and negative signs of the real number. The resulting flux phase $\phi$ in Eq.(\ref{phasephi}) is trivial, either $0$ or $\pi$, trivial because these phase angles do not break time-reversal symmetry. The corresponding phase angle $\psi$ from Eq.(\ref{phasepsi}) is also trivial, because $\psi_{ij}$ is always zero.

		Next, we consider the contribution of the transverse part $l_xs_x+l_ys_y$ of the transfer type spin-orbit coupling term. From Eqs.(\ref{z2overlapfirst}) and (\ref{lastparttransverse}), it is noted that $_j\langle \alpha,s|(l_xs_x+l_ys_y)|\alpha',s'\rangle_i$ is complex in general; it can be real or imaginary valued. As such, the resulting phase angles $\phi$ and $\psi$ are nontrivial; they are neither $0$ nor $\pi$, and thus break time-reversal symmetry. This explains our main conclusion that the transverse part of transfer type spin-orbit coupling induces the quantum anomalous Hall effect.
		
		Due to the topological character of the effective flux, it is independent of the strength of the spin-orbit coupling; the flux manifests once the spin-orbit coupling onset to nonzero value. The $C=1$ Chern plateau should thus be independent of the spin-orbit coupling; only its position in Fermi energy and width depends on the spin-orbit coupling $\lambda$. We have verified this numerically by computing the Chern plateau for two different values of spin-orbit coupling $\lambda$ while all other parameters are fixed.

		\section{Contribution of On-site Spin-Orbit Coupling to Quantum Anomalous Hall Effect}
		
\begin{figure}
			\includegraphics[angle=0,origin=c, scale=1.0]{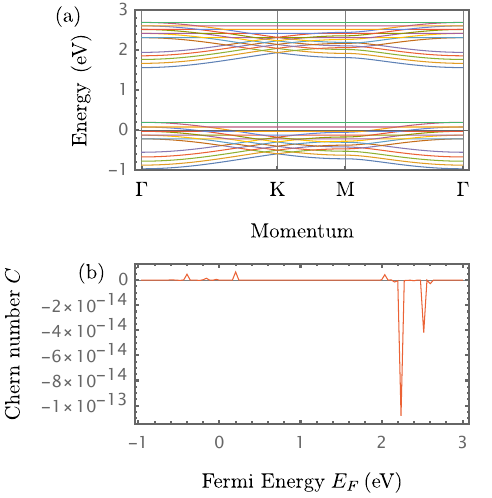}
			\caption{
				(a) Full band structure along the $K-\Gamma-M-\Gamma$ in the first Brillouin zone for the tight-binding model Eq.(\ref{Hamiltonian}) of the kagome metal-organic-framework with $E_1=E_2=E_3=1.0$eV, $V_{dd\pi}=V_{dd\delta}=V_{dd\sigma}=-0.25$eV with $\lambda_{OS}=0.2$eV, while $\lambda=0.0, M_z=2.5, M_x=M_y=0.0$ all in eV. (b) Corresponding Chern number vs. Fermi energy.}\label{fig:OnsiteSOCcontribution}
		\end{figure}

		To complete the analysis of our model, an on-site SOC turns out to be unable to generate any quantum anomalous Hall effect in the case of isotropic Slater-Koster integrals; the topological invariant is practically zero at all energies, regardless of the strength of the on-site spin-orbit interaction, as illustrated in Fig. \ref{fig:OnsiteSOCcontribution}. This is simply explained by the fact that in the isotropic limit of Slater-Koster integrals, the onsite SOC does not generate the flux pattern necessary for the QAHE as represented in Fig.\ref{fig:FluxPattern} (with mathematical proof provided detailed later in this section) and does not open up gaps at the Dirac and QBC points. As evident in Fig.\ref{fig:OnsiteSOCcontribution}, the band structure simply involves vertical shift of the bands of different orbitals, giving no gap necessary for a Chern insulating state, as described in more details in Section VIII. In other words, in the isotropic limit, the onsite SOC becomes irrelevant and its role is taken over by the transfer SOC which opens up gaps, thanks to the flux it generates.
		
		In tight-binding model, spin-orbit coupling is normally written as coupling between spin and the orbital angular momentum carried by electron hopping between nearest or next-neighbour sites. Intuitively speaking, such hopping spin-orbit coupling term effectively corresponds to flux 
		or complex phase factor acquired by the electron wave function when traversing a closed trajectory as it hops between the sites of the lattice. However, one can also consider local spin-orbit interaction, coupling the spin of electron and its local angular momentum defined on each site, as represented by the fourth term in Eq.(\ref{Hamiltonian}). The onsite spin-orbit coupling corresponds to spin-dependent uniform magnetic field in real space and induces Landau level-type of gaps, shifting bands of different orbitals away from each other, as illustrated in Fig. \ref{fig:OnsiteSOCcontribution}. Therefore, the onsite spin-orbit coupling acts like an effective Zeeman field and behaves similarly to the Zeeman-like exchange-field term (the third term; i.e., the second line in equation (\ref{Hamiltonian}), which is also locally defined on each site), which simply shifts the spin up and down states by relative amount proportional to $|\mathbf{M}|$, as reflected in the full band structure in Fig.2(a). This can be seen by noting that one can write $\lambda_{OS}\mathbf{l}\cdot\mathbf{s}=-\mathbf{M}'\cdot\mathbf{s}$, where $\mathbf{M}'=-\lambda_{OS}\mathbf{l}$, which is applicable because both the Zeeman-like exchange field and onsite spin-orbit coupling terms are onsite. Since the eigenvalues of angular momentum operator are quantized; $l_z=m_l\hbar$ where $m_l=0,\pm 1, \pm 2$ for $l=2$ ($d$-orbital), the bands for each spin sector are split into five separate bands with equal splitting proportional to $\lambda_{OS}$, as illustrated in Fig.\ref{fig:OnsiteSOCcontribution}. Such an effective field $\mathbf{M}'$ does not open up gap at the Dirac point, nor at the quadratic gapless point. In fact, other than the Zeeman field gap due to $M_z$, the band structure contains no nontrivial gap as the bands of each spin sector still overlap with each other, at least for realistic small spin-orbit coupling $\lambda_{OS}\lesssim 0.5$eV. As a result, the Chern number is practically zero and there is no quantum anomalous Hall effect due to onsite spin-orbit coupling, as confirmed in Fig. \ref{fig:OnsiteSOCcontribution}. This is an exact result in the isotropic limit of Slater-Koster integrals.
		
		It will be explained here why our model in its isotropic Slater-Koster integrals limit does not accommodate QAHE from onsite spin-orbit coupling, using flux argument, following that presented in existing works \cite{OhgushiPRB}\cite{KontaniJPSJ762007main}\cite{KontaniPRL2008main}.
		Consider a triangle in a kagome lattice. The main contribution to the flux is provided by hopping between nearest-neighbour sites, combined with the complex phase factor provided by the overlap between $d$-orbitals of similar types; $d_{xz}-d_{yz}$ orbitals, and $d_{xy}-d_{x^2-y^2}$ orbitals. These pairs of $d$-orbitals give rise to the same complex factor for spin up electrons, as given in Eqs.(\ref{overlapfirst}-\ref{overlaplast}), except that the overlap occurs at the site $j$ after the electron hops from the site $i$ to $j$. 
		
\begin{figure}
				\includegraphics[angle=0,origin=c, scale=1.0]{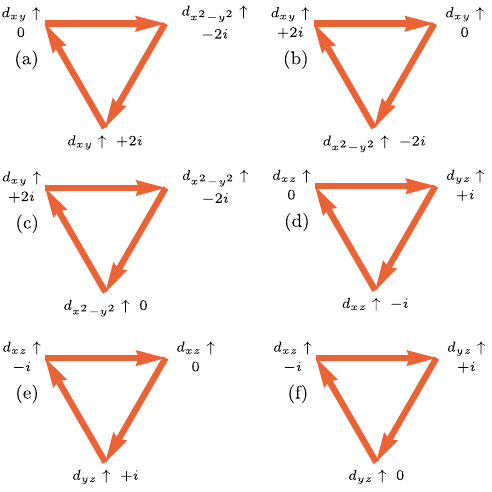}
				\caption{
					(a)-(f)The flux pattern in a ``down triangle'' of the kagome lattice due to the onsite spin-orbit coupling and nearest-neighbour hopping of electrons with orbitals $d_{\alpha}(d_{\alpha'})$ at site $i(j)$ respectively  in the low energy (occupied) spin up sector. The generally complex number attached to the end of each arrow pointing from site $i$ to site $j$ comes from the overlap factor $\langle d_{\alpha'}|l_z| d_{\alpha'}\rangle$ evaluated at site $j$. Different from the flux patterns in Figs.\ref{fig:FluxPatternExchangeSOC} and \ref{fig:FluxPatternExchangeTransverseSOCwithZ2orbital}, here the hopping between sites, as marked by the arrow, is induced by the kinetic hopping term, $t_{ij;\alpha\alpha'}$ which is non-directional; no sign change when the arrow is reversed. As such, onsite spin-orbit coupling does not induce the flux needed to open up a gap, and hence it does not generate quantum anomalous Hall effect, an exact result in the isotropic limit of Slater-Koster integrals $V_{dd\sigma}=V_{dd\delta}=V_{dd\pi}$.} \label{fig:FluxPatternOnsiteSOC}
			\end{figure}

		The resulting flux patterns are similar to those arising due to exchange-type spin-orbit coupling, as illustrated in Fig.  \ref{fig:FluxPatternOnsiteSOC}. The flux factor associated with an arrow that goes from site $i$ to $j$ is that due to onsite spin-orbit coupling at $j$ from two $d$-orbital states; between the orbital $d_{\alpha}$ at site $i$ and the orbital $d_{\alpha'}$ at site $j$. It can be seen that the net flux factor is zero in all the patterns shown in Fig.  \ref{fig:FluxPatternOnsiteSOC}. 
		
		Analogous figures of flux pattern for the ``up triangle''. Considering the flux pattern at up and down triangles, one can deduce the flux pattern for the hexagon, assuming a flux pattern where all down triangles have identical flux pattern among themselves and all up triangles have identical flux pattern among themselves. As in the exchange-type spin-orbit coupling case, the flux through the hexagons is also zero. It can therefore be concluded that the onsite spin-coupling gives zero flux at triangles and hexagons. By comparison with the kagome flux phase that hosts a quantum Hall effect \cite{OhgushiPRB}, the onsite spin-orbit coupling in our model in its isotropic Slater-Koster integrals limit thus does not  give rise to a quantum anomalous Hall effect.    
		It will be proven analytically mathematically here that this peculiar result on the absence of contribution from onsite spin-orbit coupling to quantum anomalous Hall effect is a manifestation of emergent triangular $C_3$ point group symmetry in the isotropic limit of the Slater-Koster integrals, where the perfect flatband is present. For onsite spin-orbit coupling, the phase angle $\phi_{ij}$ in the hopping from site $i$ to $j$ is given by
		\begin{equation}\label{phasephiOnsiteSOC}
			\exp(i\phi_{ij})=\frac{_j\langle \alpha,s|l_as_a|\alpha',s'\rangle_i}{|_j\langle \alpha,s|l_as_a|\alpha',s'\rangle_i|}
		\end{equation}
		where the hopping itself is induced by the kinetic hopping terms while the overlap between the two wave functions is induced by onsite spin-orbit coupling. 
		
		Now, let us consider the Ising part $l_z s_z$ of the spin-orbit coupling. As noted in the previous section, $_j\langle \alpha,s|l_z|\alpha',s'\rangle_i$ is either zero or purely imaginary. As such, $\phi_{ij}=\pm\pi/2$. Looking at Fig. \ref{fig:FluxPatternExchangeSOC}, it is clear that the phase angles $\phi_{ij}$'s always come in pair of $\pm\pi/2$ while the third one is zero. As such, the resulting phase angle $\phi$ is always zero. On the other hand, the phase angle $\psi$ becomes
		\begin{equation}\label{phasepsiOnsiteSOC}
			\psi=\sum_{ij\in\triangle,\triangledown}\psi_{ij}  \; , \; \; \;\;\tan(\psi_{ij})=\frac{\pm|\delta t_{ij}|}{t_{ij}},
		\end{equation} 
		which in general gives a nontrivial value for $\psi$; $\psi\neq 0,\pi$. However, in the isotropic limit of the Slater-Koster integrals, $t_{ij}$ is equal for all the three nearest-neighbour vectors $\mathbf{r}_{ij}$; $t_{ij}=t_0$. As a result, in this isotropic limit, 
		the resulting flux phase $\psi$ becomes trivial; $\psi=0$ because two contributions of magnitude $|\psi_{ij}|=\arctan(|\delta t_{ij}|/t_0)$ carrying opposite $\pm$ signs always come in pair when evaluating $\psi$ around a triangle.

		The transverse part of onsite spin-orbit coupling induces generally complex-valued $_j\langle \alpha,s|l_x s_x+l_ys_y|\alpha',s'\rangle_i$, as noted in previous section. The resulting phase angles $\phi$ and $\psi$ for a general set of kinetic hopping integrals $t_{ij}$'s therefore tend to be non-trivial. However, an interesting situation occurs for the case when the kinetic hopping integrals $t_{ij}$ is equal for the three nearest-neighbour vectors $\mathbf{r}_{ij}$; $t_{ij}=t_0$, manifesting a triangular ($C_3$ point group) symmetry. Then, 
		\begin{equation}\label{phasepsiOnsiteSOC}
			\psi=\sum_{ij\in\triangle,\triangledown}\psi_{ij}  \; , \; \; \;\;\tan(\psi_{ij})=\frac{|\delta t_{ij}|\sin\phi_{ij}}{t_0+|\delta t_{ij}|\cos\phi_{ij}}.
		\end{equation}
		Now, let us look at the triangles in Fig.\ref{fig:FluxPatternExchangeTransverseSOCwithZ2orbital}. Consider the following: for each triangle, one fixes the state of the electron at one of the sites while swapping the states of the electron on the other two sites. As we trace the hopping of the electron on the same (clockwise) sense, starting and ending at the site with the fixed state, since the $t_{ij}=t_0$ is equal for all the three nearest-neighbour hopping vectors, the resulting flux simply gets reversed in sign $\psi\rightarrow -\psi$, due to the fact that the roles of of the states $\alpha,s$ and $\alpha',s'$ are reversed in Eq.(\ref{phasephiOnsiteSOC}). However, since the starting and end points of the hopping are the same; at the site with the fixed state, the two contributions; $\psi$ and $-\psi$ should be summed up. As a result, the net phase angle of the flux is zero. In conclusion, in the isotropic Slater-Koster integrals limit where $t_{ij}=t_0$ is equal for all the three nearest-neighbour hopping vectors, onsite spin-orbit coupling does not induce any quantum anomalous Hall effect.

		\begin{figure*}
			\includegraphics[angle=0,origin=c, scale=2.0]{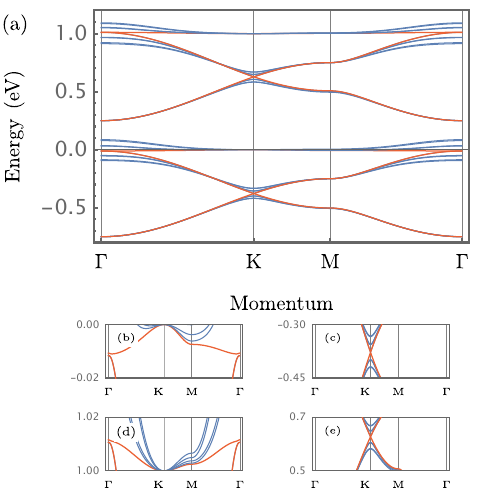}
			\caption{
				(a) Full band structure of $d$-orbital electrons  along the $K-\Gamma-M-\Gamma$ path in the first Brillouin zone (Fig.\ref{fig:KagomeLatticeandBZ}) for the kagome model Eq.(\ref{Hamiltonian}) and its projection onto the range of energy around the (b) QBC and (c) Dirac gaps of the spin up sector and (d) QBC and (e) Dirac gaps of the spin down sector, with ~\cite{Note} $E_1=E_2=E_3=0.25$eV, $V_{dd\pi}=V_{dd\delta}=V_{dd\sigma}=-0.25$eV,   $\lambda=0.05$eV and $\lambda_{OS}=0.0$eV, and $M_z=1.0,M_x=M_y=0.0$ all in eV. The color convention follows that of Fig. 2(c) and (e): red is for $d_{z^2}$ orbital while blue is for all other $d$ orbitals.}
			\label{fig:BandstructureWeakerSOC}
		\end{figure*}
		
		\section{Weaker Spin - Orbit Coupling Case}
		In the results presented in the previous sections, we have used rather large value of spin-orbit coupling $\lambda=190, 200$meV in order to highlight the possible Chern insulating states that exist in $d$-electron kagome systems. As can be seen in Fig.\ref{fig:DensityofStates}, these values of transfer spin-orbit coupling produce a gap of about 30 meV at QBC point and slightly less than 10 meV at Dirac point. While the assumed spin-orbit coupling $\lambda$ is rather strong, it is the resulting gaps that can be directly measured or compared with the result from experiment or first-principle \textit{ab-initio} calculation. Interestingly, \textit{ab-initio} calculation predicts Dirac gap of about 10 meV for the family of TM$_3$C$_6$O$_6$ compounds that we consider (See e.g. Fig. 11 in \cite{Denawi} ). Therefore, the strength of the spin-orbit coupling $\lambda$ that we use, since it is a parameter of the model, is justified, because it produces insulating gaps of the right order of magnitude.    
		
		Now, we have demonstrated in Section V that the physical mechanism that drives the Chern insulating state is the effective flux that is induced by the transfer spin-orbit coupling. This effective flux does not depend on the actual strength of the spin-orbit coupling $\lambda$, as long as it is nonzero; the effective flux immediately kicks in once the transfer spin-orbit coupling strength $\lambda$ is nonzero. As such, our results hold even for much smaller transfer spin-orbit coupling $\lambda$. In Fig. \ref{fig:BandstructureWeakerSOC}, we present the full band structure for weaker spin-orbit coupling ($\lambda=50$meV taken for example) and its projection on the spin up and spin down sectors, in energy ranges around the Dirac and QBC gaps for each sector. The figure demonstrate that the band structure has the same qualitative profile as the one with larger $\lambda$ that we have assumed so far, as shown in Fig.\ref{fig:DensityofStates}, except that the splittings of the initially degenerate bands are now much weaker. The resulting Dirac gap is much smaller, less than 1 meV, By the topological effective flux argument of Section VII, this weak spin-orbit coupling case will also produce Chern insulating states from the $d_{z^2}$ orbital, with Chern peaks at the QBC and Dirac gaps, but with much smaller bandwidths. 
		
		\section{Discussion and Conclusions}
		We have demonstrated a mechanism for the emergence of the quantum anomalous Hall effect (QAHE) in kagome  $d$-systems due to the transverse part of the  exchange-type intrinsic spin-orbit coupling (SOC),  due to an emergent extra spatial $C_3$ point group symmetry in the tight-binding model without spin-orbit coupling in the isotropic limit of Slater-Koster integrals, which is applicable to \textit{any} kagome $d$-electron system displaying a flatband in its band structure (in the absence of spin-orbit interactions). The phenomenon involves $d$-orbital electrons with $m_l=0$ ($d_{z^2}$ orbital) and the main contribution arises from the non-relativistic spectrum at the quadratic band crossing (QBC) point at the center of the Brillouin zone, instead of the usual relativistic Dirac spectrum at the corners of the Brillouin zone ~\cite{SCZhangPRL,Nature2018} or semi-Dirac spectrum ~\cite{Vanderbilt}. The spectrum is gapped by the \textit{transverse} part, rather than Ising part of the exchange-type SOC. All the other four $d$-orbitals give no QAHE, a scenario quite distinct and unexpected compared to existing works where normally these more ``nontrivial'' $d$-orbitals are considered to be responsible for the anomalous Hall effect ~\cite{KontaniPRB2007,SCZhangPRL}.
		The QAHE in the presence of exchange-type SOC has contributions both due to the QBC and Dirac points of the $d_{z^2}$ orbital,
		but the bandwidth of Chern plateau in the former is significantly broader than the latter, 
		in principle making the QAHE due to the QBC easier to achieve and detect as compared to the QAHE associated to the Dirac point. The demonstrated QAHE requires lower amount of doping in the case of the QBC point, thus offering an alternative route toward the realization of dissipationless electronic devices.

	\begin{widetext}
		
		\centering
		\textbf{Supplementary Materials:  Quantum Anomalous Hall Effect in $d$-Electron Kagome Systems: Chern Insulating States from Transverse Spin-Orbit Coupling}\\
		\centering
		I. Makhfudz$^{1}$, M. Cherkasskii$^{2}$, M. Alipourzadeh$^3$,  Y. Hajati$^3$, P. Lombardo$^1$, S. Schäfer$^1$, S. V. Kusminskiy $^{2,4}$, and R. Hayn$^{1}$
		
			\centering
		$^{1}$IM2NP, UMR CNRS 7334, Aix-Marseille Universit\'{e}, 13013 Marseille, France\\
	
	$^{2}$Institut für Theoretische Festkörperphysik, RWTH Aachen University, 52056 Aachen, Germany\\
	
	$^{3}$Department of Physics, Faculty of Science, Shahid Chamran University of Ahvaz, 6135743135 Ahvaz, Iran\\
	
	$^{4}$Max Planck Institute for the Science of Light, Staudtstraße 2, PLZ 91058 Erlangen, Germany
	\date{\today}
	\bigskip
	
	\begin{flushleft}
		In these Supplementary Materials, we provide additional details supplementing the main text. First, the basis states with respect to which the Hamiltonian matrix is written, the Slater-Koster integrals, and the diagonalization of the resulting matrix are described. Second, we provide the plots of the band structure 
		and its evolution when varying the parameter of utmost interest; the exchange-type (transfer) spin-orbit coupling. 
	\end{flushleft}
	
\end{widetext}

\section{Basis States and Diagonalization of Hamiltonian Matrix}
We focus on \emph{d}-band electrons, which have the angular momentum $l=2$ and spin $s=1/2$. 
The Hamiltonian matrix corresponding to Eq.(1) in the main text is written in the basis states involving the five $d$-orbitals, three sublattices, and two spin states (up and down spin). 

The five $d$-orbitals ($d_{xy}, d_{yz}, d_{z^2}, d_{xz}, d_{x^2-y^2}$) basis states correspond to real spherical harmonics $Y_{l,m_l}$ with $l=2$, which can be written in terms of complex spherical harmonics, on which the orbital angular momentum operator $\mathbf{l}$ operates; the complex spherical harmonics correspond to the eigenfunctions of the orbital angular momentum $l_z|l, m_l\rangle = m_l\hbar |l, m_l\rangle$, $Y^{m_l}_l(x,y,z) \equiv \langle \theta,\phi| l,m_l\rangle$. The relations are given by
\begin{equation}\label{dxy}
d_{xy}\equiv Y_{2,-2}=i\sqrt{\frac{1}{2}}\left(Y^{-2}_2-Y^{2}_2\right)
\end{equation}
\begin{equation}
d_{yz}\equiv Y_{2,-1}=i\sqrt{\frac{1}{2}}\left(Y^{-1}_2+Y^{1}_2\right)
\end{equation}
\begin{equation}
d_{z^2}\equiv Y_{2,0}=Y^0_2
\end{equation}
\begin{equation}
d_{xz}\equiv Y_{2,1}=\sqrt{\frac{1}{2}}\left(Y^{-1}_2-Y^{1}_2\right)
\end{equation}
\begin{equation}\label{dx2-y^2}
d_{x^2-y^2}\equiv Y_{2,2}=\sqrt{\frac{1}{2}}\left(Y^{-2}_2+Y^{2}_2\right)
\end{equation}
where $Y_{l,m_l}$ and $Y^{m_l}_l$ are real and complex spherical harmonics functions respectively. More generally, we can write
\begin{equation}\label{realcomplexharmonics}
|d_{\alpha}\rangle\equiv Y_{2,m_l}=\sum_{m_l}c_{\alpha;m_l}Y^{m_l}_2\equiv \sum_{m_l}c_{\alpha;m_l}|2, m_l\rangle
\end{equation}
where $\alpha=xy,yz,z^2,xz,x^2-y^2$ while $c_{\alpha;m_l}$ is complex coefficient, which can be deduced from Eqs.(\ref{dxy}-\ref{dx2-y^2}).

First, we address the kinetic hopping term. The hopping constant $t_{ij,\alpha,\alpha'}$ depends on the orbitals $d_{\alpha}(d_{\alpha'})$ at the sites $i(j)$ respectively and is determined from Slater-Koster integrals. We write the nearest neighbor vectors as 
\begin{equation}
\mathbf{r}_{ij}=a\left(l \hat{\mathbf{x}}+m\hat{\mathbf{y}}+ n\hat{\mathbf{z}}\right),
\end{equation}
where $a$ is the lattice spacing, $\hat{\mathbf{x}},\hat{\mathbf{y}},\hat{\mathbf{z}}$ are unit vectors in $x, y, z$ directions respectively. Eventually, we automatically have $n=0$ in our two-dimensional system. The Slater-Koster hopping integrals are given as follows \cite{HarrisonS};
\begin{widetext}
\begin{equation}
	t_{xz; xz}=3 n^2 l^2 V_{dd\sigma} + (n^2 + l^2 - 
	4 n^2 l^2) V_{dd\pi} + (m^2 + n^2 l^2) V_{dd\delta}
\end{equation}
\begin{equation}
	t_{xz,yz}= 3 n^2 l m V_{dd\sigma} + l m (1 - 4 n^2) V_{dd\pi} + 
	l m (n^2 - 1) V_{dd\delta}
\end{equation}
\begin{equation}
	t_{yz,yz}=3 m^2 n^2 V_{dd\sigma} + (m^2 + n^2 - 
	4 m^2 n^2) V_{dd\pi} + (l^2 + m^2 n^2) V_{dd\delta}
\end{equation}
\begin{equation}
	t_{xy,xy}=3 l^2 m^2 V_{dd\sigma} + (l^2 + m^2 - 
	4 l^2 m^2) V_{dd\pi} + (n^2 + l^2 m^2) V_{dd\delta}
\end{equation}
\begin{equation}
	t_{xy, x^2-y^2}=\frac{3}{2} l m (l^2 - m^2) V_{dd\sigma} + 
	2 l m (m^2 - l^2) V_{dd\pi} + 
	\frac{1}{2} l m (l^2 - m^2) V_{dd\delta}
\end{equation}
\begin{equation}
	t_{xy,z^2}=\sqrt{3} l m (n^2 - \frac{1}{2} (l^2 + m^2)) V_{dd\sigma} - 
	2 \sqrt{3}  l m n^2 V_{dd\pi} + 
	\frac{\sqrt{3}}{2} l m (1 + n^2) V_{dd\delta}
\end{equation}
\begin{equation}
	t_{x^2-y^2,x^2-y^2}=\frac{3}{4} (l^2 - m^2)^2 V_{dd\sigma} + (l^2 + 
	m^2 - (l^2 - m^2)^2) V_{dd\pi} + (n^2 + 
	\frac{1}{4} (l^2 - m^2)^2) V_{dd\delta}
\end{equation}
\begin{equation}
	t_{x^2-y^2,z^2}=\frac{\sqrt{3}}{2} (l^2 - m^2) (n^2 - \frac{1}{2} (l^2 + m^2)) V_{dd\sigma} + 
	\sqrt{3} n^2 (m^2 - l^2) V_{dd\pi} + 
	\frac{\sqrt{3}}{4} (1 + n^2) (l^2 - m^2) V_{dd\delta}
\end{equation}
\begin{equation}
	t_{z^2,z^2}=(n^2 -\frac{1}{2} (l^2 + m^2))^2 V_{dd\sigma} + 
	3 n^2 (l^2 + m^2) V_{dd\pi} +\frac{3}{4} (l^2 + m^2)^2 V_{dd\delta}
\end{equation}
\end{widetext}
where $V_{dd\sigma}, V_{dd\pi},V_{dd\delta}$ are the Slater-Koster integrals, while $l,m,n$ are the Cartesian components of the nearest-neighbor vectors. The algebraic functions of $l,m,n$ multiplying $V_{dd\sigma},V_{dd\pi},V_{dd\delta}$ correspond respectively to the functions $f_{\sigma}, f_{\pi}, f_{\delta}$ defined in the paragraph following Eq.(1) in the main text. The group of $\pi$-orbitals ($d_{xz}, d_{yz}$ orbitals) and the group of $\sigma$-orbitals ($d_{xy}, d_{x^2-y^2}, d_{z^2}$ orbitals) are separate and do not have nonzero Slater-Koster integrals between the two groups.

Next, we consider the spin-orbit coupling with the Hamiltonian in its continuum and first quantized form given by 
\[
H_{{\rm SOC}}=\lambda{\bf l}\cdot{\bf s}=\lambda\left(l_{x}s_{x}+l_{y}s_{y}+l_{z}s_{z}\right),
\]
which can be reshaped using $l_{\pm}=l_{x}\pm i\,l_{y}$, $s_{\pm}=s_{x}\pm i\,s_{y}$ as
\[
H_{{\rm SOC}}=\lambda{\bf l}\cdot{\bf s}=\lambda\left(\dfrac{l_{+}s_{-}}{2}+\dfrac{l_{-}s_{+}}{2}+l_{z}s_{z}\right).
\]
This Hamiltonian can be written in the basis of the eigenstates of the orbital and spin angular momenta 
\begin{equation}\label{complexsphericalharmonicsbasis}
\left\{ |l,m_{l};s,m_{s}\rangle\right\} ,
\end{equation}
where $m_{l}=0,\pm 1, \pm 2$ and $m_{s}=\pm 1/2$, which correspond to complex spherical harmonics. The angular momentum and spin operators obey the following equations
\[
\begin{array}{rl}
l_{z}|l,m_{l};s,m_{s}\rangle & =\hbar m_{l}|l,m_{l};s,m_{s}\rangle,\\
s_{z}|l,m_{l};s,m_{s}\rangle & =\hbar m_{s}|l,m_{l};s,m_{s}\rangle,\\
l_{\pm}|l,m_{l};s,m_{s}\rangle & =\hbar\sqrt{l\left(l+1\right)-m_{l}\left(m_{l}\pm1\right)}|l,m_{l}\pm1;s,m_{s}\rangle,\\
s_{\pm}|l,m_{l};s,m_{s}\rangle & =\hbar\sqrt{s\left(s+1\right)-m_{s}\left(m_{s}\pm1\right)}|l,m_{l};s,m_{s}\pm1\rangle,
\end{array}
\]
where 
\[
\langle\theta,\phi|l,m_{l};s,m_{s}\rangle=Y_{l}^{m_{l}}
\]
is the complex spherical harmonics. In the rest of these supplementary materials, we use $l_z$ to represent the $z$ projection of orbital angular momentum operator while $m_l$ the corresponding eigenvalue.

Thus, we can calculate the matrix elements
\begin{widetext}
\[
H_{{\rm SOC}}(l',m'_l,s',m'_s;l,m_l,s,m_s) =\lambda\langle l',m_{l}';s',m_{s}'|\dfrac{l_{+}s_{-}}{2}+\dfrac{l_{-}s_{+}}{2}+l_{z}s_{z}|l,m_{l};s,m_{s}\rangle
\]
\[
=\hbar^{2}\lambda\left[\dfrac{1}{2}\sqrt{l\left(l+1\right)-m_{l}\left(m_{l}+1\right)}\sqrt{s\left(s+1\right)-m_{s}\left(m_{s}-1\right)}\delta_{m_{l}',\left(m_{l}+1\right)}\delta_{m_{s}',\left(m_{s}-1\right)}\delta_{m'_s,-\frac{1}{2}}\delta_{m_s,+\frac{1}{2}}\right.
\]
\begin{equation}
	\left.+\dfrac{1}{2}\sqrt{l\left(l+1\right)-m_{l}\left(m_{l}-1\right)}\sqrt{s\left(s+1\right)-m_{s}\left(m_{s}+1\right)}\delta_{m_{l}',\left(m_{l}-1\right)}\delta_{m_{s}',\left(m_{s}+1\right)}\delta_{m'_s,+\frac{1}{2}}\delta_{m_s,-\frac{1}{2}}+m_{l}m_{s}\delta_{m_{l}',m_{l}}\delta_{m_{s}',m_{s}}\right].
\end{equation}
\end{widetext}
Explicit substitution of $l=2$ and $s=1/2$ results in
\begin{widetext}
\[
H_{{\rm SOC}}(l',m'_l,s',m'_s;l,m_l,s,m_s) =\hbar^{2}\lambda\left[\dfrac{1}{2}\sqrt{6-m_{l}\left(m_{l}+1\right)}\sqrt{3/4-m_{s}\left(m_{s}-1\right)}\delta_{m_{l}',\left(m_{l}+1\right)}\delta_{m_{s}',\left(m_{s}-1\right)}\delta_{m'_s,-\frac{1}{2}}\delta_{m_s,+\frac{1}{2}}\right.
\]
\begin{equation}
	\left.+\dfrac{1}{2}\sqrt{6-m_{l}\left(m_{l}-1\right)}\sqrt{3/4-m_{s}\left(m_{s}+1\right)}\delta_{m_{l}',\left(m_{l}-1\right)}\delta_{m_{s}',\left(m_{s}+1\right)}\delta_{m'_s,+\frac{1}{2}}\delta_{m_s,-\frac{1}{2}}+m_{l}m_{s}\delta_{m_{l}',m_{l}}\delta_{m_{s}',m_{s}}\right].
\end{equation}
\end{widetext}
Alternatively, the spin-orbit coupling Hamiltonian can be written in the basis of real spherical harmonics, which correspond directly to the wave functions of the $d$-orbitals 
\[
\left\{ |d_{\alpha};s,m_{s}\rangle\right\} ,
\]
where, as before, $\alpha=xy,yz,z^2,xz,x^2-y^2$. That is, we evaluate the following matrix elements,
\[
H_{\mathrm{SOC}}(\alpha, s',m'_s;\beta,s,m_s)=
\]
\[
\lambda\langle d_{\alpha};s',m_{s}'|\dfrac{l_{+}s_{-}}{2}+\dfrac{l_{-}s_{+}}{2}+l_{z}s_{z}|d_{\beta};s,m_{s}\rangle
\]
\[
= \lambda\sum_{m'_l,m_l}c^*_{\alpha, m'_l}c_{\beta, m_l}\langle l',m'_l;s',m_{s}'|\dfrac{l_{+}s_{-}}{2}+\dfrac{l_{-}s_{+}}{2}+l_{z}s_{z}|l,m_l;s,m_{s}\rangle
\]
\begin{equation}
=\lambda\sum_{m'_l,m_l}c^*_{\alpha, m'_l}c_{\beta, m_l}H_{{\rm SOC}}(l',m'_l,s',m'_s;l,m_l,s,m_s)
\end{equation}
where we have used Eq. (\ref{realcomplexharmonics}). 

The kinetic hopping part of the Hamiltonian gives rise to matrix elements that can be directly extracted from Slater-Koster integrals, as available in \cite{HarrisonS}. The kinetic hopping part does not depend on spin; it does not flip spin, and therefore only appears in the spin up-spin up and spin down-spin down sectors. That is, the matrix elements are proportional to $\delta_{m'_s,m_s}$. Since the kinetic part is evaluated in the basis of $\left\{d_{\alpha},s,m_s\right\}$, in the calculation, we have used this real spherical harmonics basis instead of the basis based on complex spherical harmonics basis Eq.(\ref{complexsphericalharmonicsbasis}). 

\section{Evolution of Band Structure with Spin-Orbit Coupling}
Spin-orbit coupling (between nearest-neighbor sites) plays critical role because it is directly responsible for quantum anomalous Hall effect. In general, spin-orbit coupling (the nearest-neighbor, i.e. the last term in Eq.(1) in the main text) opens up a gap at the gapless points. In general, in the absence of spin-orbit coupling, the gapless points can occur at high symmetry points such as the $\Gamma$ point at the center of the Brillouin zone as well as the $K$ point at the corner of the Brillouin zone, as well as at low symmetry points away from the center or the corners of the hexagonal Brillouin zone. For the $s$-like band structure, due to the isotropic Slater-Koster integrals $V_{dd\pi}=V_{dd\sigma}=V_{dd\delta}$, as is the case in our calculation, the gapless points occur only at high-symmetry spots.  

\begin{figure}[h!]
\includegraphics[angle=0,origin=c, scale=0.4]{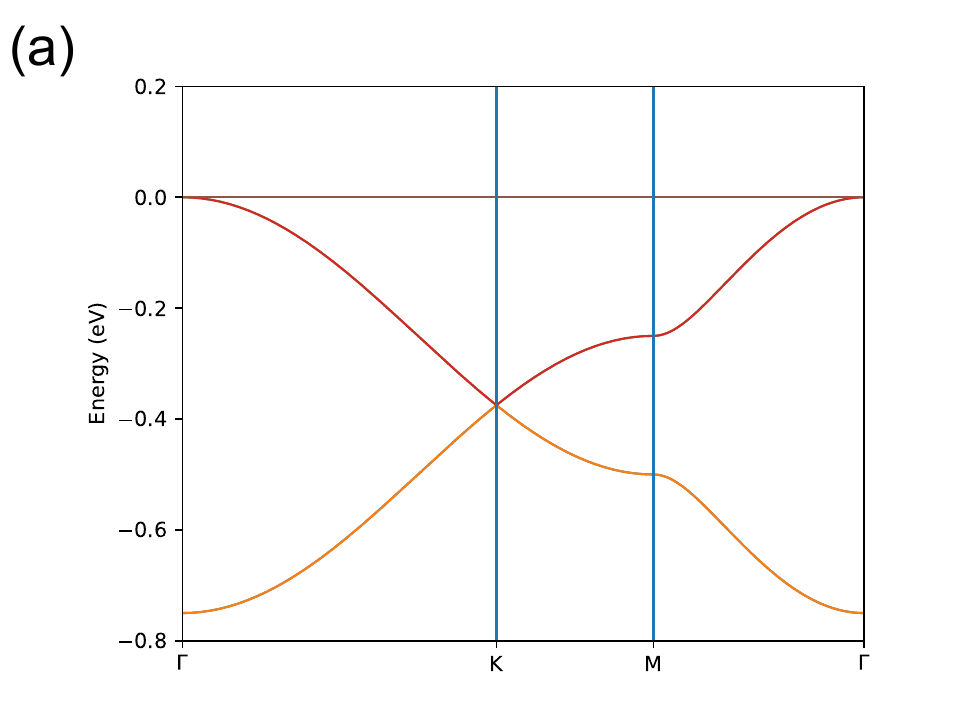}
\includegraphics[angle=0,origin=c, scale=0.4]{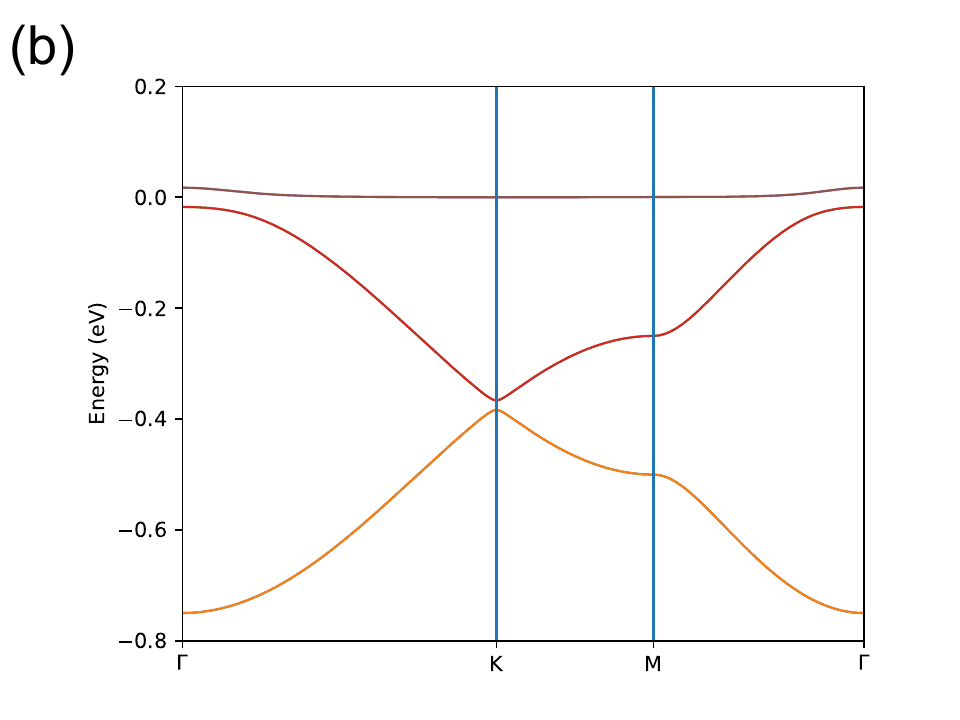}
\includegraphics[angle=0,origin=c, scale=0.4]{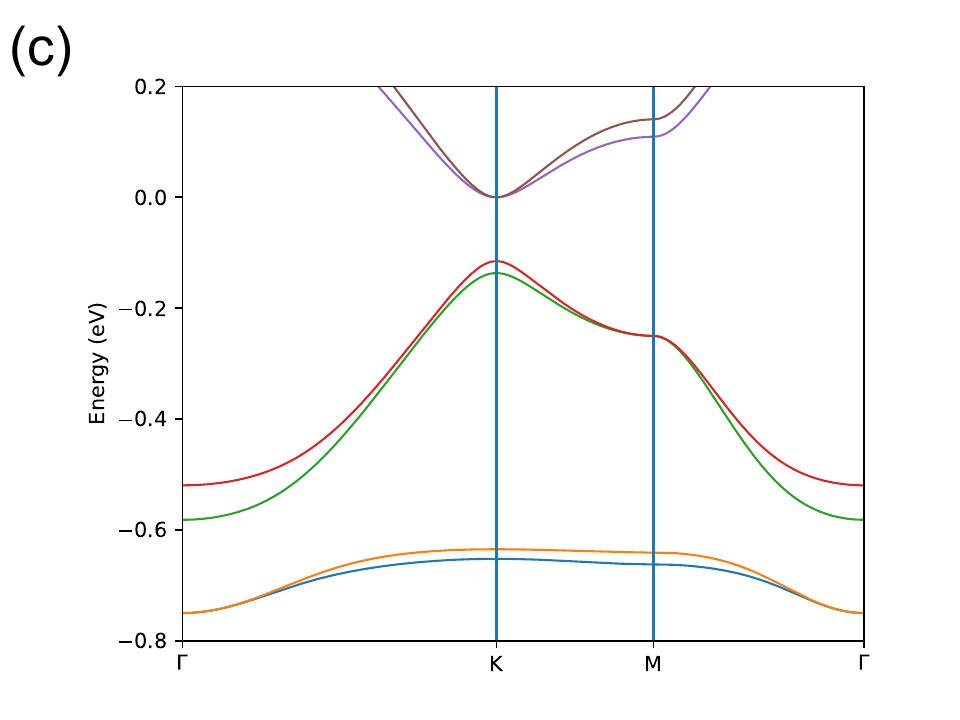}
\caption{
	The evolution of the band structure of the (degenerate) $d_{xy}$ and $d_{x^2-y^2}$ orbitals along the $K-\Gamma-M-\Gamma$ in the first Brillouin zone for the tight-binding model Eq.(1) in the main text of the kagome metal-organic-framework  with $E_1=E_2=2.0$eV, $E_3=1.0$eV, $V_{dd\pi}=V_{dd\delta}=V_{dd\sigma}=-0.25$eV in the spin up sector with a) $\lambda=0.0$eV, b) $\lambda=0.01$eV, and c) $\lambda=0.3$eV while $\lambda_{OS}=0,M_z=2.5,M_x=M_y=0.0$ all in eV. A ``band inversion'' phenomenon is visible.}
\label{fig:BandStructureXYX2Y2}
\end{figure}

We demonstrate here the occurrence of the remarkable phenomenon of band inversion, as one increases the spin-orbit coupling. Fig. ~\ref{fig:BandStructureXYX2Y2} shows the band structure in the lower energy regime, with parameters chosen to correspond to $d_{xy}$ and $d_{x^2-y^2}$ orbitals, in the spin up sector.

We observe that the flat band gets inverted: for weak $\lambda$, the flat band is on top of two dispersive bands. However, for stronger $\lambda$ ($\gtrsim 0.3$eV in the example), the flat band falls below the dispersive bands. This can be viewed as a ``band inversion'' phenomenon. It is to be noted that for these two orbitals $d_{xy}$ and $d_{x^2-y^2}$, the spin-orbit coupling opens up a rather significant gap at both the Dirac point at $K$ and the quadratic band crossing at $\Gamma$, due to the fact that these orbitals have finite $m_l=\pm 2$.  

\begin{figure}
\includegraphics[angle=0,origin=c, scale=0.4]{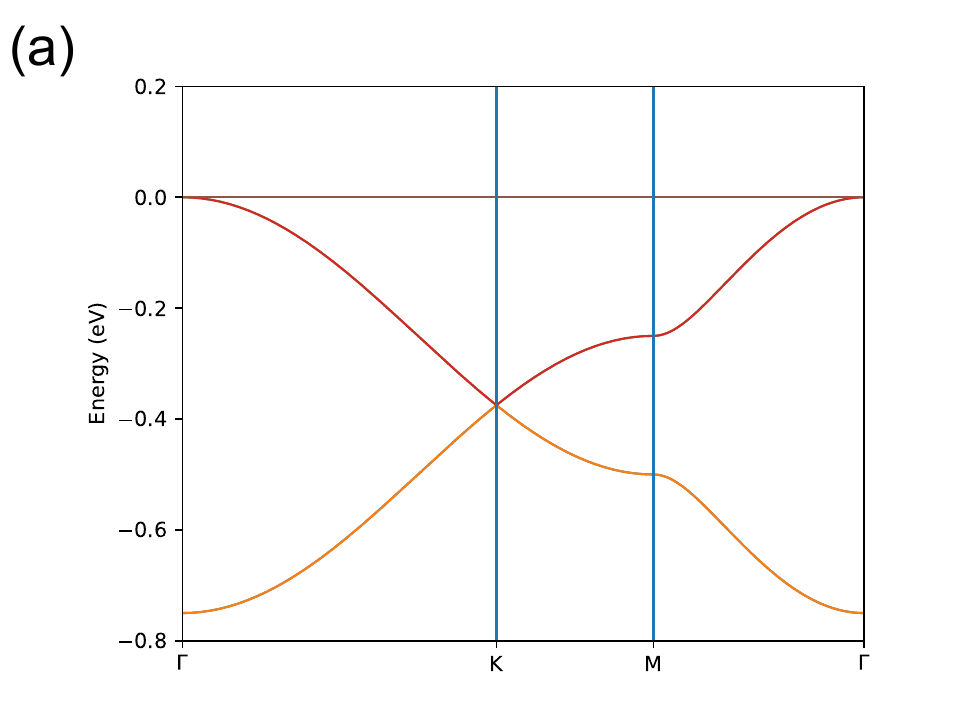}
\includegraphics[angle=0,origin=c, scale=0.4]{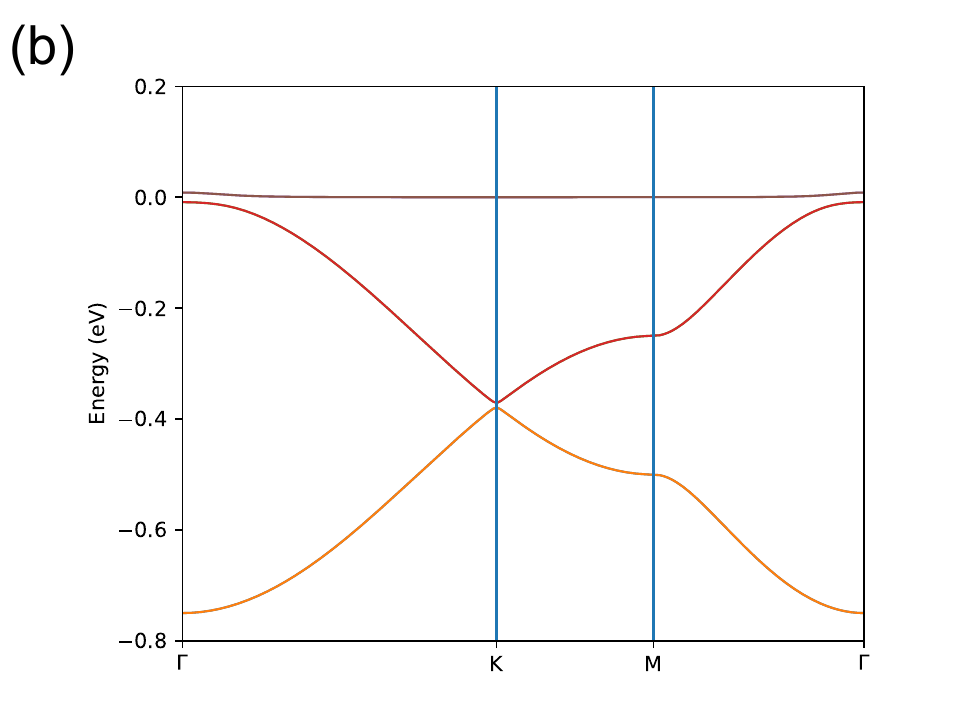}
\includegraphics[angle=0,origin=c, scale=0.4]{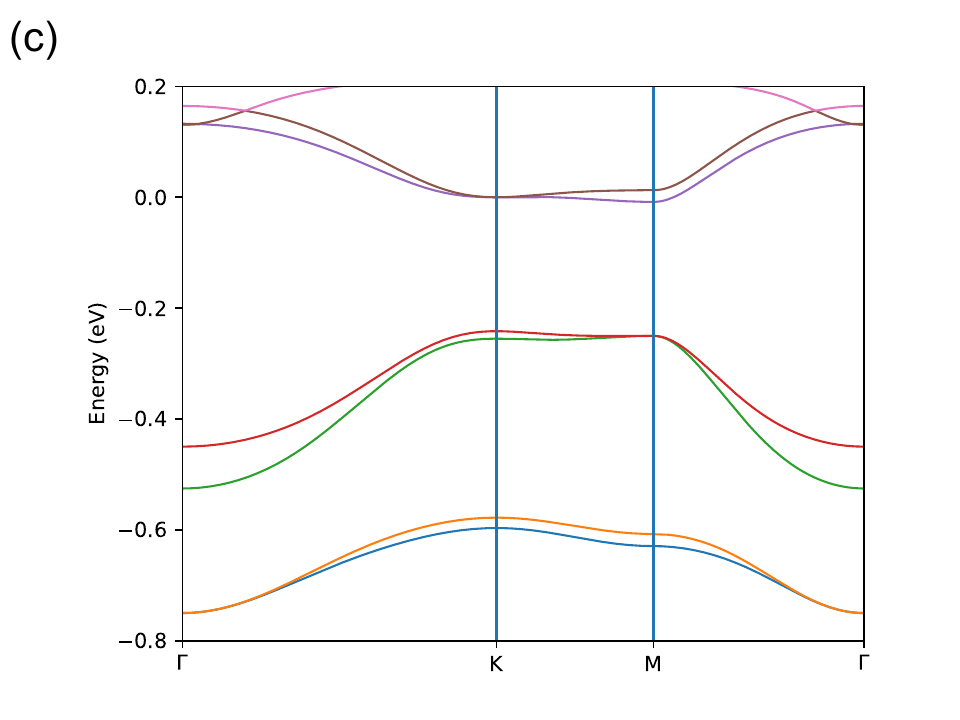}
\caption{
	The evolution of the band structure of the (degenerate) $d_{xz}$ and $d_{yz}$ orbitals along the $K-\Gamma-M-\Gamma$ in the first Brillouin zone for the tight-binding model Eq.(1) in the main text of the kagome metal-organic-framework with $E_1=E_3=2.0$eV, $E_2=1.0$eV, $V_{dd\pi}=V_{dd\delta}=V_{dd\sigma}=-0.25$eV in the spin up sector with a) $\lambda=0.0$eV, b) $\lambda=0.01$eV, and c) $\lambda=0.4$eV while $\lambda_{OS}=0.0,M_z=2.5,M_x=M_y=0.0$ all in eV. A similar ``band inversion'' phenomenon is visible.} \label{fig:BandStructureXZYZ}
\end{figure}

Similar phenomenon of band inversion occurs to $d_{xz}$ and $d_{yz}$ orbitals. At weak spin-orbit coupling $\lambda$, the flat band occurs at the top while at strong $\lambda$, the flat band occurs at the bottom, as illustrated in Fig. \ref{fig:BandStructureXZYZ}. The gaps opened at Dirac and quadratic band crossing points are also reasonably significant, due to $m_l=\pm 1$.

\begin{figure}
	\includegraphics[angle=0,origin=c, scale=0.4]{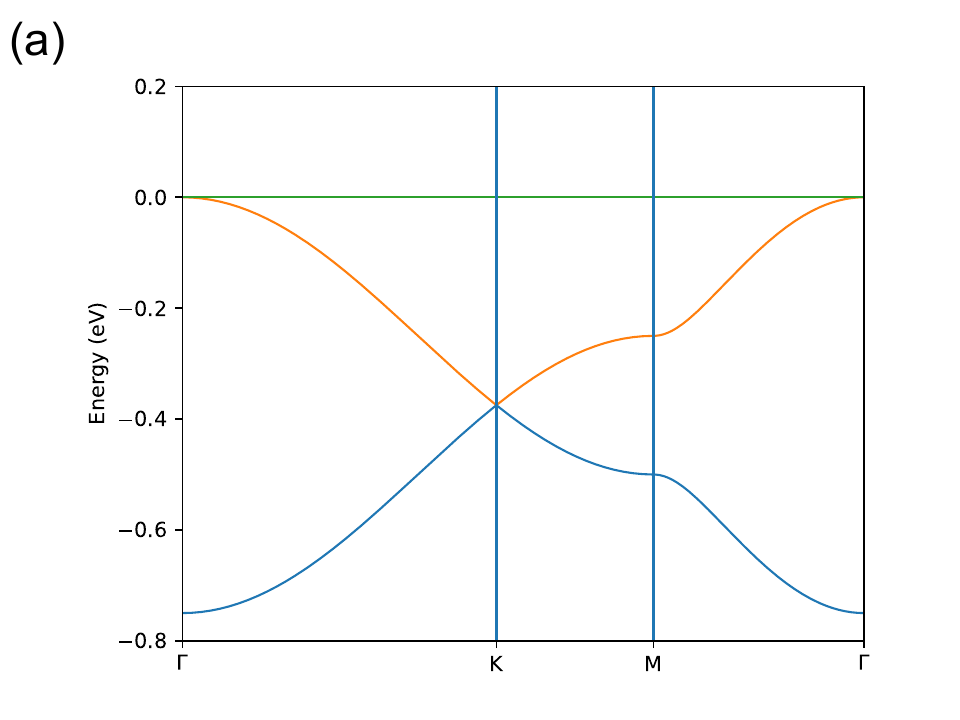}
	\includegraphics[angle=0,origin=c, scale=0.4]{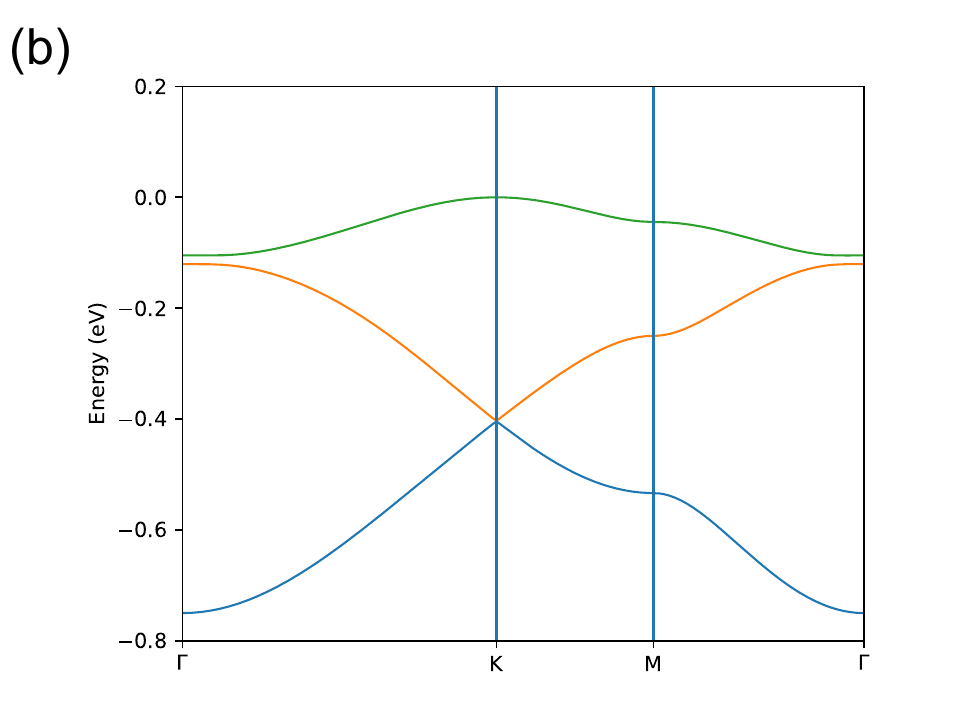}
	\includegraphics[angle=0,origin=c, scale=0.4]{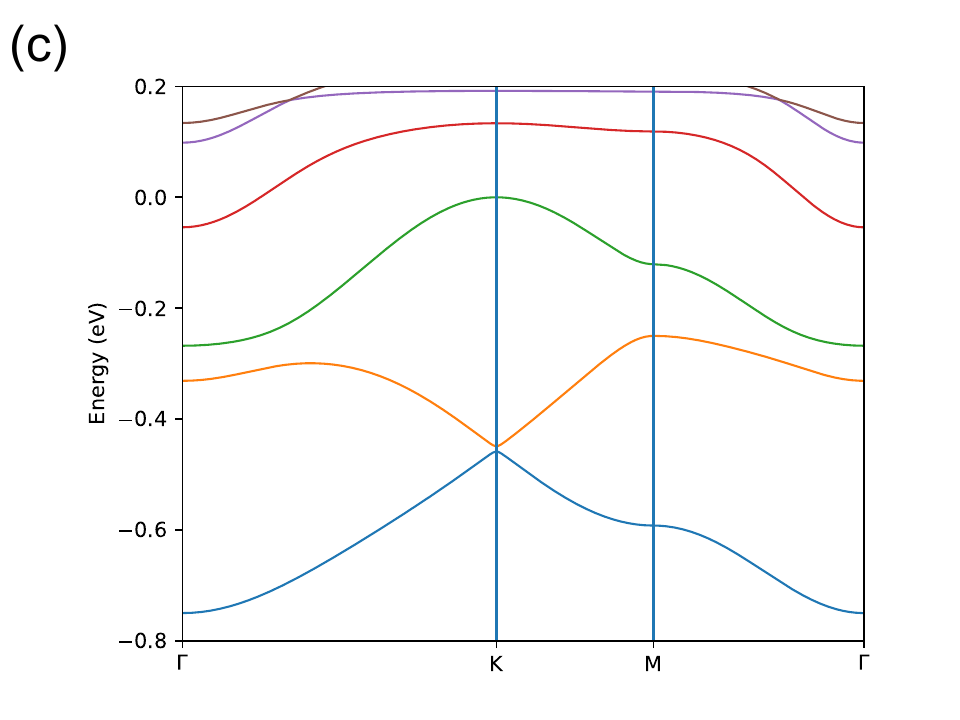}
	\caption{
		The evolution of the band structure of the (degenerate) $d_{z^2}$ orbital along the $K-\Gamma-M-\Gamma$ in the first Brillouin zone for the tight-binding model Eq.(1) in the main text of the kagome metal-organic-framework with $E_1=1.0$eV, $E_2=E_3=2.0$eV, $V_{dd\pi}=V_{dd\delta}=V_{dd\sigma}=-0.25$eV in the spin up sector with a) $\lambda=0.0$eV (with gapless Dirac point), b) $\lambda=0.3$eV (with extremely small gap at the Dirac point), and c) $\lambda=0.5$eV, while $\lambda_{OS}=0,M_z=2.5,M_x=M_y=0.0$ all in eV. No band inversion occurs. The Dirac point at $K$ acquires a very small gap while the quadratic band crossing at $\Gamma$ opens up a significant gap, as visible in c).}\label{fig:BandStructureZ2}
\end{figure}

On the other hand, for $d_{z^2}$ orbital, this band inversion phenomenon does not occur. In addition, because $m_l=0$ in this case, the Dirac point only acquires very small gap due to the $l_xs_x+l_ys_y$ term in the spin-orbit coupling. On the other hand, quadratic brand crossing acquires a gap that is larger by one order of magnitude or more than that at the Dirac point, as illustrated in Fig. \ref{fig:BandStructureZ2}.

\begin{thebibliography}{1}
			
			\bibitem{AHErmp}N. Nagaosa, J. Sinova, S. Onoda, A. H. MacDonald, and N. P. Ong, \textit{Anomalous Hall effect}, Rev. Mod. Phys. 82, 1539 (2010).
			
			\bibitem{KarplusLuttinger}R. Karplus and J. M. Luttinger, \textit{Hall Effect in Ferromagnetics}, Phys. Rev. 95, 1154
			(1954).
			
			\bibitem{Smit}J. Smit, \textit{The spontaneous hall effect in ferromagnetics I}, Physica (Amsterdam) 21,  877 (1955).
			
			\bibitem{Luttinger}J. M. Luttinger, \textit{Theory of the Hall Effect in Ferromagnetic Substances}, Phys. Rev. 112, 739 (1958).
			
			\bibitem{Kondo}J. Kondo, \textit{Anomalous Hall Effect and Magnetoresistance of Ferromagnetic Metals}, Prog. Theor. Phys. 27, 772 (1962).
			
			\bibitem{Berger}L. Berger, \textit{Side-Jump Mechanism for the Hall Effect of Ferromagnets}, Phys. Rev. B 2, 4559 (1970).
			
			\bibitem{Nozieres}P. Nozieres and C. Lewiner, \textit{A simple theory of the anomalous hall effect in semiconductors}, J. Phys. (Paris) 34, 901
			(1973).
			
			\bibitem{JungwirthPRL}T. Jungwirth, Qian Niu, and A. H. MacDonald, \textit{Anomalous Hall Effect in Ferromagnetic Semiconductors}, Phys. Rev. Lett. 88, 207208 (2002). 
			
			\bibitem{OnodaNagaosaJPSJ}M. Onoda and N. Nogaosa, \textit{Topological Nature of Anomalous Hall Effect in Ferromagnets}, J. Phys. Soc. Jpn. 71, 19 (2002).
			
			\bibitem{KontaniPRB2007}H. Kontani, T. Tanaka, and K. Yamada, \textit{Intrinsic anomalous Hall effect in ferromagnetic metals studied by the multi-$d$-orbital tight-binding model},
			Phys. Rev. B 75, 184416 (2007).
			
			\bibitem{Flatband1}T. Neupert, L. Santos, C. Chamon, and C. Mudry, \textit{Fractional Quantum Hall States at Zero Magnetic Field}, Phys. Rev. Lett. 106, 236804 (2011).
			
			\bibitem{Flatband2}K. Sun, Z. Gu, H. Katsura, and S. Das Sarma, \textit{Nearly Flatbands with Nontrivial Topology},
			Phys. Rev. Lett. 106, 236803 (2011).
			
			\bibitem{Flatband3}E. Tang, J-W. Mei, and X-G. Wen, \textit{High-Temperature Fractional Quantum Hall States},
			Phys. Rev. Lett. 106, 236802 (2011).
			
			\bibitem{DiracFlatbandKagome}M. Kang, L. Ye, S. Fang, J-S. You, A. Levitan, M. Han, J. I. Facio, C. Jozwiak, A. Bostwick, E. Rotenberg, M. K. Chan, R. D. McDonald, D. Graf, K. Kaznatcheev, E. Vescovo, D. C. Bell, E. Kaxiras, J. van den Brink, M. Richter, M. P. Ghimire, J. G. Checkelsky, and R. Comin, \textit{Dirac fermions and flat bands in the ideal kagome metal FeSn}, Nature Materials 19, 163-169 (2020).
			
			\bibitem{Zhang}R. Zhang, J. Liu, Y. Gao, M. Hua, B. Xia, P. Knecht, A. C. Papageorgiou, J. Reichert, J.
			V Barth, H. Xu, L. Huang, and N. Lin, \textit{On-surface Synthesis of a Semiconducting 2D Metal–Organic Framework Cu$_3$(C$_6$O$_6$) Exhibiting Dispersive Electronic Bands}, Angew. Chem. Int. Ed. 10, 13698 (2019).
			
			\bibitem{Hua}M. Hua, B. Xia, M. Wang, E. Li, J. Liu, T. Wu, Y. Wang, R. Li, H. Ding, J. Hu, Y.
			Wang, J. Zhu, H. Xu, W. Zhao, and N. Lin, \textit{Highly Degenerate Ground States in a Frustrated Antiferromagnetic Kagome Lattice in a Two-Dimensional Metal–Organic Framework}, J. Phys. Chem. Lett. 12, 3733 (2021).
			
			\bibitem{Shaiek}N. Shaiek, H. Denawi, M. Koudia, R. Hayn, S. Schäfer, I. Berbezier, C. Lamine, O. Siri,
			A. Akremi, and M. Abel, \textit{Self-Organized Kagomé-Lattice in a Conductive Metal-Organic Monolayer}, Advanced Materials Interfaces 9, 2201099 (2022).
			
			\bibitem{Denawi}A. H. Denawi, X. Bouju, M. Abel, J. Richter, and R. Hayn, \textit{Metal-organic kagome systems as candidates to study spin liquids, spin ice or the quantum anomalous Hall effect}, Phys. Rev. Materials 7, 074201 (2023).
			
			\bibitem{MakhfudzPRB2014}I. Makhfudz, \textit{Fluctuation-induced first-order quantum phase transition of the $U$(1) spin liquid in a pyrochlore quantum spin ice}, Phys. Rev. B 89, 024401 (2014).
			
			\bibitem{SavaryBalentsRPP}L. Savary and L. Balents, \textit{Quantum spin liquids: a review}, Rep. Prog. Phys. 80, 016502 (2017).
			
			\bibitem{NatCommsHotta}S. Nishimoto, N. Shibata, and C. Hotta, \textit{Controlling frustrated liquids and solids with an applied field in a kagome Heisenberg antiferromagnet},
			Nat. Commun. 4, 2287 (2013).
			
			\bibitem{CapponiPRB2013}S. Capponi, O. Derzhko, A. Honecker, A. M. Läuchli, and J. Richter, \textit{Numerical study of magnetization plateaus in the spin-12 kagome Heisenberg antiferromagnet},
			Phys. Rev. B 88, 144416 (2013).
			\bibitem{MakhfudzPujolPRL2015}I. Makhfudz and P. Pujol, \textit{Hole Properties On and Off Magnetization Plateaus in 2D Antiferromagnets}, Phys. Rev. Lett. 114, 087204 (2015).
			
			\bibitem{Taguchi}Taguchi Y, Oohara Y, Yoshizawa H, Nagaosa N, and Tokura Y., \textit{Spin Chirality, Berry Phase, and Anomalous Hall Effect in a Frustrated Ferromagnet}, Science 291(5513), 2573 (2001).
			
			\bibitem{YePRL}J. Ye, Y. B. Kim, A. J. Millis, B. I. Shraiman, P. Majumdar, and Z. Tešanović, \textit{Berry Phase Theory of the Anomalous Hall Effect: Application to Colossal Magnetoresistance Manganites}, Phys. Rev. Lett. 83, 3737 (1999).
			
			
			\bibitem{OnodaNagaosaPRL2}S. Onoda and N. Nagaosa, \textit{Spin Chirality Fluctuations and Anomalous Hall Effect in Itinerant Ferromagnets
			}, Phys. Rev. Lett. 90, 196602 (2003).
			
			\bibitem{MartinBatista}I. Martin and C. D. Batista, \textit{Itinerant Electron-Driven Chiral Magnetic Ordering and Spontaneous Quantum Hall Effect in Triangular Lattice Models}, Phys. Rev. Lett. 101, 156402 (2008).
			
			\bibitem{OhgushiPRB}Ohgushi K., Murakami S., Nagaosa N., \textit{Spin anisotropy and quantum Hall effect in the kagomé lattice: Chiral spin state based on a ferromagnet}, Phys. Rev. B 62, R6065 (2000).
			
			\bibitem{ZFangScience}Z. Fang, N. Nagaosa, K. S. Takahashi, A. Asamitsu, R. Mathieu, T. Ogasawara, H. Yamada, M. Kawasaki, Y. Tokura, and K. Terakura, \textit{The Anomalous Hall Effect and Magnetic Monopoles in Momentum Space}, Science 302, 92 (2003).
			
			\bibitem{CanalsPRB}M. Taillefumier, B. Canals, C. Lacroix, V. K. Dugaev, and P. Bruno,    \textit{Anomalous Hall effect due to spin chirality in the Kagomé lattice},
			Phys. Rev. B 74, 085105 (2006).
			
			\bibitem{MacDonaldPRL}H. Chen, Q. Niu, and A. H. MacDonald, \textit{Anomalous Hall Effect Arising from Noncollinear Antiferromagnetism}, Phys. Rev. Lett. 112, 017205 (2014).
			
			\bibitem{BrunoPRL}P. Bruno, V. K. Dugaev, and M. Taillefumier, \textit{Topological Hall Effect and Berry Phase in Magnetic Nanostructures}, Phys. Rev. Lett. 93, 096806 (2004).
			
			\bibitem{MakhfudzPujolPRB2015}I. Makhfudz and P. Pujol, \textit{Protection against a spin gap in two-dimensional insulating antiferromagnets with a Chern-Simons term}, Phys. Rev. B 92, 144507 (2015).
			
			\bibitem{SundaramNiu}G. Sundaram and Q. Niu, \textit{Wave-packet dynamics in slowly perturbed crystals: Gradient corrections and Berry-phase effects}, Phys. Rev. B 59, 14 915 (1999).
			
			\bibitem{Sinitsyn}N. A. Sinitsyn, \textit{Semiclassical theories of the anomalous Hall effect}, J. Phys.; Cond. Matt. 20 (2008) 023201.
			
			\bibitem{Haldane2004PRL}F. D. M. Haldane, \textit{Berry Curvature on the Fermi Surface: Anomalous Hall Effect as a Topological Fermi-Liquid Property}, Phys. Rev. Lett. 93, 206602 (2004).
			
			\bibitem{TKNN}D. J. Thouless, M. Kohmoto, M. P. Nightingale, and M. den Nijs, \textit{Quantized Hall Conductance in a Two-Dimensional Periodic Potential}, Phys. Rev. Lett. 49, 405 (1982).
			
			
			\bibitem{MacDonaldRMP}C-Z. Chang, C-X. Liu, and A. H. MacDonald, \textit{Colloquium: Quantum anomalous Hall effect},
			Rev. Mod. Phys. 95, 011002 (2023).
			
			\bibitem{Haldane1988PRL}F. D. M. Haldane, \textit{Model for a Quantum Hall Effect without Landau Levels: Condensed-Matter Realization of the "Parity Anomaly"},
			Phys. Rev. Lett. 61, 2015 (1988).
			
			\bibitem{OnodaNagaosaPRL1}M. Onoda and N. Nagaosa, \textit{Quantized Anomalous Hall Effect in Two-Dimensional Ferromagnets: Quantum Hall Effect in Metals}, Phys. Rev. Lett. 90, 206601 (2003).
			
			\bibitem{KSunPRL2009}K. Sun, H. Yao, E. Fradkin, and S. A. Kivelson, \textit{Topological Insulators and Nematic Phases from Spontaneous Symmetry Breaking in 2D Fermi Systems with a Quadratic Band Crossing},
			Phys. Rev. Lett. 103, 046811 (2009).
			
			\bibitem{QAHEgrapheneAFM}Z.Qiao, W. Ren, H. Chen, L. Bellaiche, Z. Zhang, A. H. MacDonald, and Q. Niu, \textit{Quantum Anomalous Hall Effect in Graphene Proximity Coupled to an Antiferromagnetic Insulator},
			Phys. Rev. Lett. 112, 116404 (2014).  
			
			\bibitem{QAHEinGraphene}P. Högl, T. Frank, K. Zollner, D. Kochan, M. Gmitra, and J. Fabian, \textit{Quantum Anomalous Hall Effects in Graphene from Proximity-Induced Uniform and Staggered Spin-Orbit and Exchange Coupling}, Phys. Rev. Lett. 124, 136403 (2020).
			
			\bibitem{CXLiuPRL}C-X. Liu, X-L. Qi, X. Dai, Z. Fang, and S-C. Zhang, \textit{Quantum Anomalous Hall Effect in Hg$_{1-y}$Mn$_y$Te Quantum Wells}, Phys. Rev. Lett. 101, 146802 (2008).
			
			\bibitem{CXLiuAnnCondMat}C-X. Liu, S-C. Zhang, and X-L. Qi, \textit{The Quantum Anomalous Hall Effect: Theory and Experiment}, Annu. Rev. Condens. Matter Phys. 7, 301 (2016).
			
			\bibitem{Science}C-Z. Chang, J. Zhang, X. Feng, J. Shen, Z. Zhang, M. Guo, K. Li, Y. Ou, P. Wei, L-L. Wang, Z-Q. Ji, Y. Feng, S. Ji, X. Chen, J. Jia, X. Dai, Z. Fang, S-C. Zhang, K. He, Y. Wang, L. Lu, X-C. Ma, and Q-K. Xue, \textit{Experimental Observation of the Quantum Anomalous Hall Effect in a Magnetic Topological Insulator}, Science 340, 167 (2013).
			
			\bibitem{ScienceDeng}Deng, Y., Yu, Y., Shi, M. Z., Guo, Z., Xu, Z., Wang, J., Chen, X.
			H., and Zhang, Y., \textit{Quantum anomalous Hall effect in intrinsic magnetic topological insulator MnBi$_2$Te$_4$}, Science 367, 895 (2020).
			
			\bibitem{Vanderbilt}H. Huang, Z. Liu, H. Zhang, W. Duan, and D. Vanderbilt, \textit{Emergence of a Chern-insulating state from a semi-Dirac dispersion}, Phys. Rev. B 92, 161115(R)(2015).
			
			\bibitem{QAHE2DorganicTI}Z. F. Wang, Z. Liu, and F. Liu, \textit{Quantum Anomalous Hall Effect in 2D Organic Topological Insulators}, Phys. Rev. Lett. 110, 196801 (2013).
			
			\bibitem{PRLabinitiMOF}Z. Liu, Z-F. Wang, J-W. Mei, Y-S. Wu, and F. Liu, \textit{Flat Chern Band in a Two-Dimensional Organometallic Framework}, Phys. Rev. Lett. 110, 106804 (2013).
			
			\bibitem{SciRep}S. Baidya, S. Kang, C. H. Kim, and J. Yu, \textit{Chern insulator with a nearly flat band in the metal-organic-framework-based Kagome lattice}, Scientific Reports 9, 13807 (2019).
			
			\bibitem{SCZhangPRL}G. Xu, B. Lian, and S-C. Zhang, \textit{Intrinsic Quantum Anomalous Hall Effect in the Kagome Lattice Cs$_2$LiMn$_3$F$_{12}$}, Phys. Rev. Lett. 115, 186802 (2015).
			
			\bibitem{Nature2018}L. Ye, M. Kang, J. Liu, F. von Cube, C. R. Wicker, T. Suzuki, C. Jozwiak, A. Bostwick, E. Rotenberg, D. C. Bell, L. Fu, R. Comin, and J. G. Checkelsky, \textit{Massive Dirac fermions in a ferromagnetic kagome metal}, Nature 555, 638 (2018).
			
			\bibitem{Okamoto}S. Okamoto, N. Mohanta, E. Dagotto, and D. N. Sheng, \textit{Topological flat bands in a kagome lattice multiorbital system},
			Communications Physics 5, 198 (2022).
			
			\bibitem{Harrison}W. A. Harrison, \textit{Electronic Structure and the Properties of Solids: The Physics of the Chemical Bond} (Dover Publications, Inc., New York, United States (1980)).
			
			\bibitem{SupplementaryMaterials} Please see the Supplementary Materials containing the diagonalization of the Hamiltonian including the details of Slater-Koster integrals and the evolution of band structure with transfer (exchange type) spin-orbit coupling.  The supplementary materials also include references to \cite{AHErmp},\cite{KontaniPRB2007},\cite{Denawi},\cite{OhgushiPRB},\cite{Haldane1988PRL},\cite{SCZhangPRL}, \cite{Harrison},\cite{Fukui},\cite{KontaniJPSJ762007main}, and \cite{KontaniPRL2008main}.
			
			\bibitem{KaneMelePRL}C. L. Kane and E. J. Mele, \textit{Quantum Spin Hall Effect in Graphene},
			Phys. Rev. Lett. 95, 226801 (2005).
			
			\bibitem{Note}As discussed later in Section VIII, the onsite spin-orbit coupling $\lambda_{OS}$ turns out to be complementary; i.e. unable to produce any QAHE in the isotropic Slater-Koster integrals limit of our interest, and is thus set to zero for simplicity. The in-plane magnetization $(M_x, M_y)$ is also set to zero as we assume magnetization perpendicular to the plane.
			
			\bibitem{Fukui}T. Fukui, Y. Hatsugai, and H. Suzuki, \textit{Chern Numbers in Discretized Brillouin Zone: Efficient Method of Computing (Spin) Hall Conductances}, J. Phys. Soc. Jpn. 74, 61675 (2005).
			
			\bibitem{KontaniJPSJ762007main}H. Kontani, M. Naito, D. S. Hirashima, K. Yamada, and J. Inoue, \textit{Study of Intrinsic Spin and Orbital Hall Effects in Pt Based on a (6s, 6p, 5d) Tight-Binding Model}, J. Phys. Soc. Jpn. 76, 103702 (2007).
			
			\bibitem{KontaniPRL2008main}H. Kontani, T. Tanaka, D. S. Hirashima, K. Yamada, and J. Inoue, \textit{Giant Intrinsic Spin and Orbital Hall Effects in Sr$_2$MO$_4$(M=Ru, Rh, Mo)},
			Phys. Rev. Lett. 100, 096601 (2008).
			
		\end{thebibliography}

\begin{thebibliography}{1}
	
	\bibitem{HarrisonS}W. A. Harrison, \textit{Electronic Structure and the Properties of Solids: The Physics of the Chemical Bond} (Dover Publications, Inc., New York (1980)).
	
	\bibitem{AHErmpS}N. Nagaosa, J. Sinova, S. Onoda, A. H. MacDonald, and N. P. Ong, Rev. Mod. Phys. 82, 1539 (2010).
	
	\bibitem{FukuiS}T. Fukui, Y. Hatsugai, and H. Suzuki, J. Phys. Soc. Jpn. 74, 61675 (2005).
	
	\bibitem{Haldane1988PRLs}F. D. M. Haldane,
	Phys. Rev. Lett. 61, 2015 (1988).
	
	\bibitem{OhgushiPRBs}Ohgushi K., Murakami S., Nagaosa N., Phys. Rev. B 62, R6065 (2000).
	
	\bibitem{DenawiS}A. H. Denawi, X. Bouju, M. Abel, J. Richter, and R. Hayn, Phys. Rev. Materials 7, 074201 (2023).
	
	\bibitem{KontaniPRB2007SM}H. Kontani, T. Tanaka, and K. Yamada,
	Phys. Rev. B 75, 184416 (2007).
	
	\bibitem{SCZhangPRL2015SM}G. Xu, B. Lian, and S-C. Zhang, Phys. Rev. Lett. 115, 186802 (2015).
	
	\bibitem{KontaniJPSJ762007}H. Kontani, M. Naito, D. S. Hirashima, K. Yamada, and J. Inoue, J. Phys. Soc. Jpn. 76, 103702 (2007).
	
	\bibitem{KontaniPRL2008}H. Kontani, T. Tanaka, D. S. Hirashima, K. Yamada, and J. Inoue,
	Phys. Rev. Lett. 100, 096601 (2008).
	
\end{thebibliography}
\end{document}